\documentclass[a4paper,10pt,twocolumn]{article}
\pdfoutput=1
\usepackage{XCharter}
\usepackage[T1]{fontenc}
\usepackage{blkarray}
\usepackage{booktabs}
\usepackage{array}
\usepackage{fullpage}
\usepackage{amsfonts}
\usepackage{amsmath}
\usepackage{slashed}
\usepackage{amssymb}
\usepackage{graphicx}
\usepackage{orcidlink}
\usepackage[utf8]{inputenc}
\usepackage[normalem]{ulem}
\usepackage{latexsym}
\usepackage[dvipsnames]{xcolor}
\usepackage[export]{adjustbox}
\usepackage{float}
\usepackage{multirow}

\usepackage{enumitem}
\usepackage{lipsum}
\usepackage{hyperref}
\hypersetup{colorlinks=true,citecolor=purple,linkcolor=NavyBlue,urlcolor=NavyBlue}
\usepackage{tabularx}
\usepackage[numbers,sort&compress]{natbib}
\usepackage{relsize}
\usepackage[left=1.6cm,right=1.6cm,top=2.0cm,bottom=1.8cm]{geometry}
\usepackage{comment}
\usepackage{shorthand}
\usepackage{dcolumn}
\usepackage{multirow}
\usepackage{booktabs}

\usepackage{longtable}
\usepackage{siunitx}

\usepackage{pifont}
\begin{document}

\title{Bayesian Inferences on Analytical Equations of State Approximations of Neutron Stars}

\author{%
Arijit Das\orcidlink{0000-0002-3116-0571}\\
\small Indian Institute of Science Education and Research Thiruvananthapuram\\
\small Vithura, Kerala 695551, India\\
\small \texttt{arijit21@iisertvm.ac.in}
\\[1.2em]
Sourav Roy Chowdhury\orcidlink{0000-0003-2802-4138}\\
\small Vidyasagar College, 39 Sankar Ghose Lane\\
\small Kolkata 700006, India\\
\small \texttt{souravrc79@gmail.com}
}
\date{}

\maketitle

\begin{abstract}
Equations of state (EoS) for dense matter are commonly provided in tabulated pressure-energy density relations, complicating numerical implementation and potentially compromising thermodynamic consistency. In this work, we propose a piecewise-continuous analytical form with a common parametrization capable of representing a broad class of dense matter EoSs. We validate this parametrization against tabulated EoSs from the CompOSE and LALSimulation repositories. Following the initial deterministic fitting, we employed two distinct Bayesian approaches: one based on synthetic EoSs generated by adding noise to the fitted EoS and another based on multi-messenger measurements of neutron-star masses and tidal deformabilities. In both approaches, the initial best-fit parameters are used as reference values. For the considered EoSs, spanning from very soft to very stiff, the resulting fits reproduce the tabulated EoSs and the associated macroscopic neutron-star observables in the repositories with sufficient accuracy. The inferred parameters exhibit strong correlations arising from the continuity conditions imposed at the segment boundaries. These correlations persist in the low- and intermediate-density segments under multi-messenger inference but become weak in the core, reflecting the limited constraining power of current observations at high densities. The ability of a single functional form to describe EoSs derived from different microscopic frameworks, together with the recurrence of similar parameter correlations, suggests that diverse dense matter EoSs may share a common underlying structure. \vfill
\end{abstract} 

\section{Introduction}
Neutron stars (NSs) are among the densest known astrophysical objects composed of ultra-dense nuclear and/or quark matter core \cite{Mandal:2009uk,Mandal:2012fq,Mandal:2016dzg,Mandal:2017ihr,Das:2026usu}, making their dense cores very natural astrophysical testing ground for dense QCD, hadron physics and even exotic scenarios e.g. dark matter \cite{Das:2026fyp,Das:2025skn}, at densities far exceeding nuclear saturation density~\cite{Bombaci:2018ksa} ($\rho_0$) by several times, far beyond the reach of current terrestrial experiments~\cite{Lattimer:2012nd,Sumiyoshi2020}. However, despite significant efforts, the exact composition of matter at such extreme densities remains uncertain.
\par Identifying the correct NS EoS, the central ingredient for understanding the macroscopic structure of NS structure, its evolution, core-collapse supernovae, and compact binary mergers~\cite{Douchin:2001sv,Bombaci:2018ksa,Sumiyoshi2020,Huth:2021bsp}, remains a fundamental challenge in nuclear and particle astrophysics. 
Depending on the density regime, various frameworks have been used to calculate the EoS. Below $\rho < 2\rho_0$, chiral effective field theory has been widely used with well-quantified uncertainties. This is complemented by experiments e.g. HADES \cite{Spies:2024zoh,HADES:2019auv}, CBM \cite{Heuser:2013bua,Senger:2020pzs,Senger:2020wvj,CBM:2016kpk,Agarwal:2022ydl}, RHIC \cite{PHENIX:2015siv,STAR:2010vob,STAR:2010vob}, MPD at NICA \cite{MPD:2022qhn,Kisiel:2021cmn,MPD:2025jzd,Vereschagin:2025dap}, PREX-II \cite{PREX:2021umo}, MREX/CREX\cite{CREX:2022kgg}, GSI \cite{KAOS:2000ekm,Fuchs:2000kp}, STAR \cite{STAR:2017sal}, PHENIX \cite{PHENIX:2015siv}, GMR \cite{Li:2007bp} and RCNP \cite{Tamii:2011pv} whose measurements of various nuclear properties \cite{Drischler:2015eba,Drischler:2017wtt,Abrahamyan:2012gp,Ding:2024xxu,PREX:2021umo,Tsang:2012se,Russotto:2016ucm,Tsang:2012se,Lattimer:2014scr,Burgio:2024vjc,Reed:2021nqk,Drischler:2015eba,Drischler:2017wtt,Dutra:2012mb}
can be used to constrain the low energy part of the EoS. On the other hand, for asymptotically high baryon densities ($\rho \geq 40\rho_0$), the EoS can be obtained from pQCD calculations quite reliably. However, at intermediate densities ($\sim 5\rho_0-8\rho_0$) relevant for NSs, nothing much about the composition of the inner crust and the core is known. Moreover, the lack of first principles methods forces the usage of either phenomenological of effective models, some of which are shown in Table.~\ref{eos_category}. Countless propositions of such models exist, including nuclear matter, confined or deconfined quarks, hyperonic matter, strange quark matter \cite{Deb:2016lvi,Chowdhury:2019wte}, etc. 
\par As terrestrial experiments have not been able to explore this density regime, efforts have instead been invested in astronomical observations which have the potential of substantially narrowing the allowed EoS parameter space~\cite{Tsang:2023vhh,Koehn:2024set}. Notable examples include the NICER mission that employs pulse profile modelling to simultaneously determine NS masses and radii~\cite{Riley:2019yda,Miller:2019cac,Riley:2021pdl,Miller:2021qha,Choudhury:2024xbk,Qi:2025mpn}, measurements of $\Lambda_{1.4}$ by LIGO-Virgo-Kagra collaboration~\cite{Ozel:2016oaf,Lattimer2021ARNPS,Raaijmakers:2021uju,Riley:2019yda,Miller:2019cac,Miller:2021qha,Choudhury:2024xbk, Most:2018hfd,Biswas:2021yge,Li:2022ozz,Most:2018hfd,Biswas:2021yge,Li:2022ozz} and NS mass measurements of pulsar timing campaigns~\cite{Fonseca:2021wxt,Romani:2022jhd}. Future probes, e.g. eXTP \cite{Watts:2018iom}, ATHENA \cite{Hauf:2011fn}, Einstein Telescope \cite{ET:2019dnz} and its American counterpart the Cosmic Explorer \cite{Evans:2021gyd} are expected to tighten the constraints even further. A list of recent measurements of such macroscopic observables is listed in Table.~\ref{tab:Data}.
In fact, the combination of NICER constraints with $\Lambda_{1.4}$ bounds has proven powerful in breaking degeneracies that neither probe can resolve independently~\cite{Raaijmakers:2021uju,Biswas:2021yge,Jiang:2019rcw}, especially with the detection of GW170817~\cite{LIGOScientific:2017vwq} showing that GW signals encode information about the NS interior during inspiral and post-merger phase. These observations have established GW astronomy as a robust probe of NS EoS~\cite{Radice:2016rys,HernandezVivanco:2019vvk}.
\par A whole body of ready made EoSs can be found in publicly available repositories, e.g. CompOSE \cite{CompOSECoreTeam:2022ddl} and LALSimulation \cite{lalsuite}, usually provided in the form of three column tables with the baryonic density $\rho$, the energy density $\mathcal{E}$, and the pressure $\mathcal{P}$. As modeling the crust of a NS is fundamentally different from the core, these repositories are abundant in nonuniform EoSs, where the crust and the core part of the EoS are not computed from the same underlying model and the crust and the core part of the EoS are stitched together. The crust-core interface is usually constructed using an \textit{ad hoc} matching procedure, which can lead to unphysical behavior at the interface. However, such EoS have been shown to lead to considerable errors in the computation of NS observables \cite{Suleiman:2021hre}.
\par Despite this sheer diversity of EoSs derived from a school of different theoretical considerations, including degrees of freedom such as nucleons, hyperons and even hybrid compositions, share strikingly similar qualitative shapes: a monotonically increasing crust, followed by smoothly varying intermediate regions with different curvatures and, an almost linearly varying core region whose stiffness is bounded by causality. This strongly suggests that independent of the microscopic origin of these EoSs, all these EoSs may have some underlying ``universal'' structure, and hence, may be described by a common, piecewise smooth and sufficiently flexible functional form, common to whole classes of EoSs. Such a universal form, if it exists, should be able to fit all these diverse families of available EoSs, from softest to the stiffest and for both unified and non-unified EoSs. Establishing whether this is indeed the case, and constructing such a parametrization explicitly, is the goal of this paper. On the other hand, even if such a universal fit is obtained, such a fit would be deterministic with no associated statistical uncertainties. Given that the multi-messenger (MM) measurements themselves have windows of uncertainties, it would be desirable to include sources of statistical uncertainties associated with the determination of the fit parameters. In this way, the parametrization will provide not only a common functional description of the diverse families of EoSs but also will account for the uncertainties in the fitted parameters and the variation among different EoSs.
\par There are considerable problems with tabular EoSs as far as numerical implementation is concerned. One needs to use such tables via interpolation, which again introduces ambiguities in the global nature of the EoS, as interpolation procedures are not unique. This can lead to ambiguities in calculations of NS observables. Moreover, interpolation should be implemented to maintain thermodynamic consistency. Moreover, from a computational perspective, using tabulated EoSs is computationally expensive. To avoid these problems, it is much better to have global analytical fits \cite{Nola:2026cvb} for these EoSs with adjustable parameters that match the tabulated EoS. Spectral fits \cite{Lindblom:2018rfr,Lindblom:2010bb} and piecewise polytropic (PP) fits \cite{Read:2008iy,Suleiman:2022egw}, and more recently, Gaussian process and neural network-based non-parametric approaches~\cite{Landry:2018prl,Essick:2019ldf,Legred:2021hdx} have already been proposed for standard unified EoSs \cite{Haensel:2004nu,Potekhin:2013qqa}.
\par In this work, we propose an alternative fitting procedure which can even be used for non-unified EoSs as well. We use this procedure to fit most of the EoSs of CompOSE and LALSimulation repositories combined. In Sec.\ref{EoS_Models}, we briefly review various theoretical approaches of dense matter EoS modeling to demonstrate the sheer expanse of the available theoretical landscapes. In section \ref{sec:eos_model}, we present the EoS parametrization used in this paper. In section \ref{sec:analytical_fit_evaluation}, we describe the numerical pathway used to obtain the analytical fitting form for each EoS and noise inclusion in the EoS to simulate uncertainties. In section \ref{sec:bayes_framework}, we introduce the Bayesian inference pipeline used to constrain the model parameters using MM measurements. In sec. \ref{sec:results}, we present the results of the fitting procedure and comment on the accuracy of fits to reproduce macroscopic properties obtained from the original EoSs. In sec. \ref{sec:correlation}, we discuss how different fitting parameters are correlated with each other, as inferred by noise-driven synthetic data (Syn) generation and MM data driven Bayesian analysis. Lastly, in sec. \ref{sec:conclusion} we present concluding remarks.


\par 

\section{Neutron Star EoS Models}
\label{EoS_Models}
The EoS is defined as the fundamental thermodynamic relation between pressure $\mathcal{P}$, energy density $\mathcal{E}$, and baryon number density $\rho$. 
Now, because the structure of a NS is stratified, i.e. different density regimes of NSs have different degrees of freedom, a unified EoS for all density regimes is fairly difficult to obtain. The low energy density regime is constrained by measurements on heavy, terrestrial nuclei, while the inner crust may harbour exotic phases of nuclear matter, e.g. gnocchi, spaghetti, lasagna, bucatini and swiss-cheese at the very boundary of inner crust and outer core. Whereas the outer core at densities $\rho_0 \lesssim \rho_b \lesssim 2 \rho_0$ is well described by chiral effective field theory ~\cite{Hebeler:2010jx,Tews:2012fj}, the inner core at $\rho \gtrsim 2\text{--}8\,n_0$ remains the most uncertain and the most speculative region where a zoo of exotic propositions exist e.g.  \cite{Zhu:2026zga,Mandal:2009uk,Mandal:2012fq,Mandal:2016dzg,Andersen:2025ezj,Pradhan:2025pol,Shahrbaf:2025hsw,Geissel:2025vnp,Mu:2026cqr,Pang:2023dqj}. A central challenge in constructing NS EoS lies in modeling matter consistently across these widely varying density regimes. In the next subsections, we briefly describe some theoretical approaches used to model the dense matter EoS.

\subsection{Microscopic Variational Many-Body Models}
In this approach, EoS is calculated directly from realistic NN and NNN interactions. As an example, the Wiringa, Fiks and Fabrocini (WFF) family \cite{Wiringa:WFF_family} employs microscopic hamiltonians of the form 
\begin{align}
    H_{\text{WFF}} = -\sum_{i}\frac{\nabla^2_i}{2m} +\sum_{i<j}v_{ij} + \sum_{i<j<k}v_{ijk}
\end{align}
where $v_{ij}$ is the NN potential fitted using NN scattering data and deuteron properties and $v_{ijk}$ is an NNN interaction. The energy density functional $\mathcal{E}(\rho)$ is calculated using the variational method with correlation operators. As an example, the WFF family of EoSs uses U14 and Argonne $v_{14}$
(A14) \cite{Wiringa:1984tg} models of the two-body potential, together with the Urbana VII (UVII) model of NNN interaction. Similarly, the APR family~\cite{Akmal:AP_family} uses the relativistic boost corrected \cite{Forest:1995sg} A18 model for NN interaction, the Urbana model IX (UIX) of NNN interactions \cite{Pudliner:1995wk}. In both cases, the NNN interactions are added by hand to reproduce the nuclear saturation data. In the Dirac--Brueckner--Hartree--Fock (DBHF) approach, self-consistently solves the in-medium G-matrix via the Bethe-Goldstone equation using the Dirac equation for nucleons and predicting a stiffer EoS. For instance, MPA1~\cite{Muther:MPA_family} is the representative of DBHF model. These approaches are usually reliable up to a few times $\rho_0$.

\subsection{Chiral Effective Field Theory}
ChEFT organizes nuclear forces as a systematic expansion in $(Q/\Lambda_\chi)^n$ (where $Q$ is the characteristic momentum transfer, $\Lambda_\chi$ is the chiral symmetry breaking scale and $n$ labels the order of expansion) and physical observables are rigorously calculated order by order with their associated uncertainties~\cite{Hebeler:2010jx,Tews:2012fj}. However, for $\rho \geq 2\rho_0$, the ChEFT expansion breaks down. 
For instance, the basic idea behind the QHC18 EoS is to separate the star into domains: the outer and the intermediate crust regions below neutron drip, the neutron drip region, the nuclear pasta region, the low and high density regions of the liquid nuclear matter, the crossover, and higher density quark matter core. Accurate polynomial fits of the numerical EoS are used
\begin{align}
    \mathcal{E}(\xi) &= a\xi + d_0 + d_1\xi \text{ln}\xi + \sum_{\nu = 2}^{\nu_\text{max}}d_\nu\xi^{\ell_\nu}\\
    \mathcal{P}(\xi) &= -d_0 + d_1\xi +\sum_{\nu = 2}^{\nu_{\text{max}}}(\ell_\nu - 1)d_\nu\xi^{\ell_{\nu}},
\end{align}
with the coefficients $a$, $\{d_i\}$ in the polynomials tuned to each domain. 
\subsection{Phenomenological Models:}
The phenomenological approach parametrizes nuclear interactions through effective degrees of freedom calibrated to finite-nucleus data. Non-relativistic Skyrme EDFs describe nucleon interactions via a zero-range density-dependent contact force. 
As an example, the SLY family~\cite{Douchin:SLYn4,Danielewicz:SLY2n9n230b,Gulminelli:SLY2n9n230a,Douchin:2001sv} uses the Skyrme effective interaction hamiltonian \cite{Chabanat:1997un}
\begin{align}
    \mathcal{H} = \mathcal{K} + \mathcal{H}_0 + \mathcal{H}_3 + \mathcal{H}_{\text{eff}} + \mathcal{H}_{\text{fin}} + \mathcal{H}_{\text{SO}} + \mathcal{H}_{\text{SG}} + \mathcal{H}_{\text{Coul}}
\end{align}
where $\mathcal{K}$ is the kinetic term, $\mathcal{H}_0$ is the zero-range term, $\mathcal{H}_3$ is the density dependent term, $\mathcal{H}_{\text{eff}}$ is the effective mass term, $\mathcal{H}_{\text{fin}}$ is the finite range term, $\mathcal{H}_{\text{SO}}$ is the spin-orbit term, $\mathcal{H}_{\text{SG}}$ is a term due to the tensor coupling with spin and gradient and $\mathcal{H}_{\text{Coul}}$ is the coulomb contribution. The details of the interaction model can be found in Ref.\cite{Chabanat:1997un}. The $\mathcal{H}_0$ term is purely phenomenological chosen for computational convenience and not derived from the underlying strong interaction. In order to construct the EoS nuclei in the crust are modelled using the
Compressible Liquid Drop Model, with parameters calculated using the many-body methods presented in Ref.\cite{Douchin:2000kad,Douchin:2000kx}. The calculation of the EoS is extrapolated to densities relevant for NS core assuming the simplest $npe\mu$ composition. Similarly, the BSK family~\cite{Goriely:bsk1,Chamel:bsk2} uses Hartree-Fock-Bogoliubov (HFB) method with phenomenological two body interactions with various mass models \cite{Goriely:2009zzb,Chamel:2009yx,Goriely:2010bm}. The parameters of the model are fitted to all known nuclear masses and anchors its high-density behavior to a target microscopic neutron matter EoS. The SKI family~\cite{Reinhard:1995zz} uses an extended spin-orbit term, spanning a range of symmetry energy slopes that translate directly to a spread in predicted NS radii.
\subsection{Relativistic Mean-Field and Hybrid Models}  
RMF models describe nuclear matter through various variations of the sigma model where nucleons interact via scalar, vector and isovector ($\vec{\rho}_\mu$) mesons. In the context of relativistic mean field models, the much used model is the non-linear chiral sigma model
\begin{align}\nonumber
    \mathcal{L} = &\overline{\psi}\gamma^\mu D_\mu\psi - g_\sigma\sigma + \frac{1}{2}\partial_{\mu}\sigma\partial^{\mu}\sigma - \frac{1}{2}m^2_\sigma\sigma^2 - \frac{1}{4}\Omega_{\mu\nu}\Omega^{\mu\nu} \\ &+ \frac{1}{2}m^2_\omega \omega_{\mu}\omega^\mu - \frac{1}{4}\vec{\rho}_{\mu\nu}\cdot\vec{\rho}^{\mu\nu} + \frac{1}{2}m^2_\rho\vec{\rho}_\mu\cdot\vec{\rho}^\mu - \mathcal{U}(\sigma)
\end{align}
Given the model one can very straightforwardly calculate the EoS presented in Ref.\cite{Glendenning:2005ix}. The parameters of the model are usually constrained by saturation properties of nuclear matter e.g. binding energy, nuclear incompressibility, symmetry energy and slope parameter, and then extrapolated to supra-nuclear densities. The Boguta and Bodmer potential term \cite{Boguta:1977xi} for $\sigma$ meson
\begin{align}
\mathcal{U}(\sigma) = -\frac{\kappa}{3!}\sigma^3 -\frac{\lambda}{4!}\sigma^4
\end{align}
is added to correctly reproduce nuclear incompressibility.
\par However, this model is too stiff far above saturation, threatening causality and too large a stellar mass, which is solved by including meson self interaction term of the form $\sim (\omega_\mu\omega^\mu)^2$ \cite{Sugahara:1993wz}, $\sim (\vec{\rho}_\mu\cdot\vec{\rho}^\mu)^2$ \cite{Mueller:1996pm} and $(\omega_{\nu}\omega^\nu)(\vec{\rho}_\mu\cdot\vec{\rho}^\mu)$.
\par The MS family of EoS~\cite{Mueller:MS1_family} serves as a classic example of RMF EoSs based on the Furnstahl-Serot-Tang (FST) extended RMF model \cite{Furnstahl:1996wv} where all possible nonlinear interactions are added in the energy density functional. Couplings related to low density terms are calibrated at measurements near $\rho_0$. However, as the couplings of high density terms e.g. $\sim(\omega_\mu \omega^\mu)^2$ and $\sim(\omega_\mu \omega^\mu)(\vec{\rho}_\mu\cdot\vec{\rho}^\mu)$ can not be probed near $\rho \approx \rho_0$, those couplings are kept free and depending on the values of those couplings, one obtains a sequence of EoSs: MS0, MS1, MS2, MS3 and MS1B which differ in the high density stiffness. They produce the same observed properties near saturation but very different maximum NS masses. The stiffest of these EoSs is the MS1 for which the terms $(\omega_\mu \omega^\mu)^2$ and $(\omega_\mu \omega^\mu)(\vec{\rho}_\mu\cdot\vec{\rho}^\mu)$ are simply switched off ($\zeta = \xi = 0$).
Density-dependent RMF variants introduce medium-modified meson-nucleon couplings \cite{Typel:1999yq} that generally soften \cite{Lalazissis:2005de} the EoS. On the other hand, EoS families e.g. GS~\cite{Glendenning:GS_family} occupy the intermediate stiffness range.
\par Beyond nucleonic matter, several EoS families introduce additional degrees of freedom at $n_b \gtrsim 2-3\,n_0$. Such EoSs incorporate strange baryons which soften the EoS and reduce the maximum mass~\cite{Bombaci:2018ksa}. This is known as the hyperon puzzle \cite{Bombaci:2016xzl}. The H family~\cite{Lackey:H_family,GLENDENNING1982392,Glendenning:1984jr,Glendenning:1991es,Knorren:1995ds} contains a host of hyperon-meson coupling. Moreover, hybrid EoSs often incorporate a first-order hadron-to-quark phase transition at some critical density $n_c \gtrsim 2\,n_0$. The ALF family~\cite{Alford:ALF_family} uses the MIT bag model for the quark phase, with the bag constant $B$ and vector coupling controlling the transition density and post-transition stiffness. The PCL family ~\cite{Prakash:PCL_family} combines an RMF hadronic phase with a bag model quark phase. To synthesize all the above information about the EoS, Table.~\ref{eos_category} summarizes the commonly used models by family, category, and theoretical basis. Broadly, they fall into microscopic (\textit{Micro}) and phenomenological (\textit{Pheno}) approaches.

\begin{table*}[htp!]
\centering
\renewcommand{\arraystretch}{0.85}
\setlength{\tabcolsep}{2pt}

\begin{tabularx}{\textwidth}{
@{}
l
c
c
>{\raggedright\arraybackslash}X
@{}
}
\toprule
\toprule
\textbf{EoS Family} &
\textbf{Category} &
\textbf{Sub-category} &
\textbf{Origin / Basis} \\
\midrule
\midrule
AP (AP1--4)~\cite{Akmal:AP_family} & Micro & Variational many-body & Realistic NN + 3N forces (Argonne + Urbana) \\
WFF (WFF1--3)~\cite{Wiringa:WFF_family} & Micro & Variational many-body & Realistic NN interactions.\\
MPA1~\cite{Muther:MPA_family} & Micro & Brueckner-type & Realistic NN interactions (Dirac-Brueckner)\\
ALF (ALF1--4)~\cite{Alford:ALF_family} & Pheno & Hybrid (hadron + quark) & RMF (hadronic) + bag-like quark model\\
SLY (SLY, 2, 4, 9, 230A)~\cite{Douchin:SLYn4,Danielewicz:SLY2n9n230b,Gulminelli:SLY2n9n230a} & Pheno & Skyrme EDF & Effective NN interaction + neutron matter constraints\\
SKI (SKI2--6)~\cite{Danielewicz:SLY2n9n230b,Gulminelli:SLY2n9n230a} & Pheno & Skyrme EDF & Effective interaction (enhanced spin-orbit)\\
BSK (BSK19--21) ~\cite{Goriely:bsk1,Chamel:bsk2} & Pheno & Skyrme EDF (semi-microscopic) & Fit to nuclear masses + microscopic constraints\\
MS1 (MS1, b, \_PP, b\_PP)~\cite{Mueller:MS1_family} & Pheno & RMF + piecewise polytrope & Relativistic mean-field; PP = fit\\
H (H1--7)~\cite{Lackey:H_family} & Pheno & RMF-type & Parameterized relativistic Lagrangian\\
FPS~\cite{Friedman:FPS_family} & Pheno & Skyrme-type EDF & Fitted effective interaction\\
ENG~\cite{Engvik:ENG_family} & Pheno & Potential-based & Effective NN interaction\\
GS~\cite{Glendenning:GS_family} & Pheno & Potential-based & G-matrix inspired interaction\\
PCL~\cite{Prakash:PCL_family} & Pheno & RMF-type hybrid EoS & Relativistic meson-exchange mean-field + bag model\\
PAL~\cite{Prakash:PAL_family} & Pheno & Parameterized energy functional &Density-dependent effective interaction\\          
\bottomrule
\bottomrule
\end{tabularx}

\caption{Classification of commonly used NS EoS by category and origin.}
\label{eos_category}
\end{table*}

\section{Modelling the Equation of State}
\label{sec:eos_model}

The analytical EoS model is described by the following parameter vector,
\begin{equation}
  \centering
  \boldsymbol{\theta}
  = \biggl\{
      \{\mathcal{E}_{\rm i}\}_{i = 1}^{3(4)},\;
      K,\;
      \alpha,\;
      \{a_i,\; b_i\}_{i = 1}^{2(3)},\;
      \{c_{s,i}^{2}\}
    \biggr\},
  \label{eq:theta}
\end{equation}
(The notation 2(3) means either 2 or 3, depending on the specific EoS) which fully specifies the pressure $\mathcal{P}(\mathcal{E}, \boldsymbol{\theta})$ as a function of energy density $\mathcal{E}$ across the three (or four, depending on the family of EoS considered) appropriate ranges of $\mathcal{E}$. For $\mathcal{E} < \mathcal{E}_1$ the crust portion of the EoS is modeled by the following functional form,
\begin{equation}
  \mathcal{P}_{\text{crust}}(\mathcal{E} < \mathcal{E}_1; \theta_{\text{crust}} = \{K, \alpha\}) = K\!\left(e^{\alpha\mathcal{E}} - 1\right),
  \label{eq:crust}
\end{equation}
with adjustable parameters $K$ and $\alpha$. This is followed by two (or three, depending on the EoS being considered) bridge segments at intermediate values of $\mathcal{E}$ having the following form 
\begin{align}
    \mathcal{P}_1(\mathcal{E}_1 < \mathcal{E} < \mathcal{E}_2; \theta_1 = \{a_1, b_1\}) &= a_1~\mathcal{E}^{\alpha_1} + b_1~\mathcal{E}^{\beta_1},\\
    \mathcal{P}_2(\mathcal{E}_2 < \mathcal{E} < \mathcal{E}_3; \theta_2 = \{a_2, b_2\}) &= a_2~\mathcal{E}^{\alpha_2} + b_2~\mathcal{E}^{\beta_2}\\
    \mathcal{P}_3(\mathcal{E}_3 < \mathcal{E} < \mathcal{E}_4; \theta_3 = \{a_3, b_3\}) &= a_3~\mathcal{E}^{\alpha_3} + b_3~\mathcal{E}^{\beta_3}
\end{align}
with fixed values of the exponents $\{ \alpha_i, \beta_i \}_{i = 1}^{3}$ and adjustable parameters $\{ a_i, b_i \}_{i = 1}^{2(3)}$. The third bridge $\mathcal{P}_3$ is not usually needed except for very occasionally e.g. for BSK24, BSK25 and BSK26. These intermediate bridges are followed by piecewise-linear segments in high-density core whose slopes are parameterized by the squared speed of sound $\{c_{s,i}^{2}\}_{i = 1}$ in each of the segments. The three matching energy densities $\{\mathcal{E}_i\}_{i = 1}^{4}$ define the boundaries between different regimes of density and are required to be strictly ordered as $\mathcal{E}_1 < \mathcal{E}_2 < \mathcal{E}_3 < \mathcal{E}_4$. Continuity of the pressure at each of the matching points is enforced. For each proposed parameter set $\boldsymbol{\theta}$ from sampling, we construct a discrete EoS table on a uniform grid on $\mathcal{E}$. Candidate EoSs with either $\mathcal{P} < 0$ or $c^2_S < 0$ for any value of $\mathcal{E}$ and hence, violating thermodynamic stability, are immediately rejected. Some of the EoSs e.g. members of the BSK family, are acausal beyond sufficiently large energy density. However, as will be discussed later, during sampling we consider Gaussian prior means for some of the $c_{s,j}^{2}$ exceeding unity to remain internally consistent with the original EoSs from the repositories.

\begin{figure}[t]
    \centering
    \includegraphics[width=\linewidth]{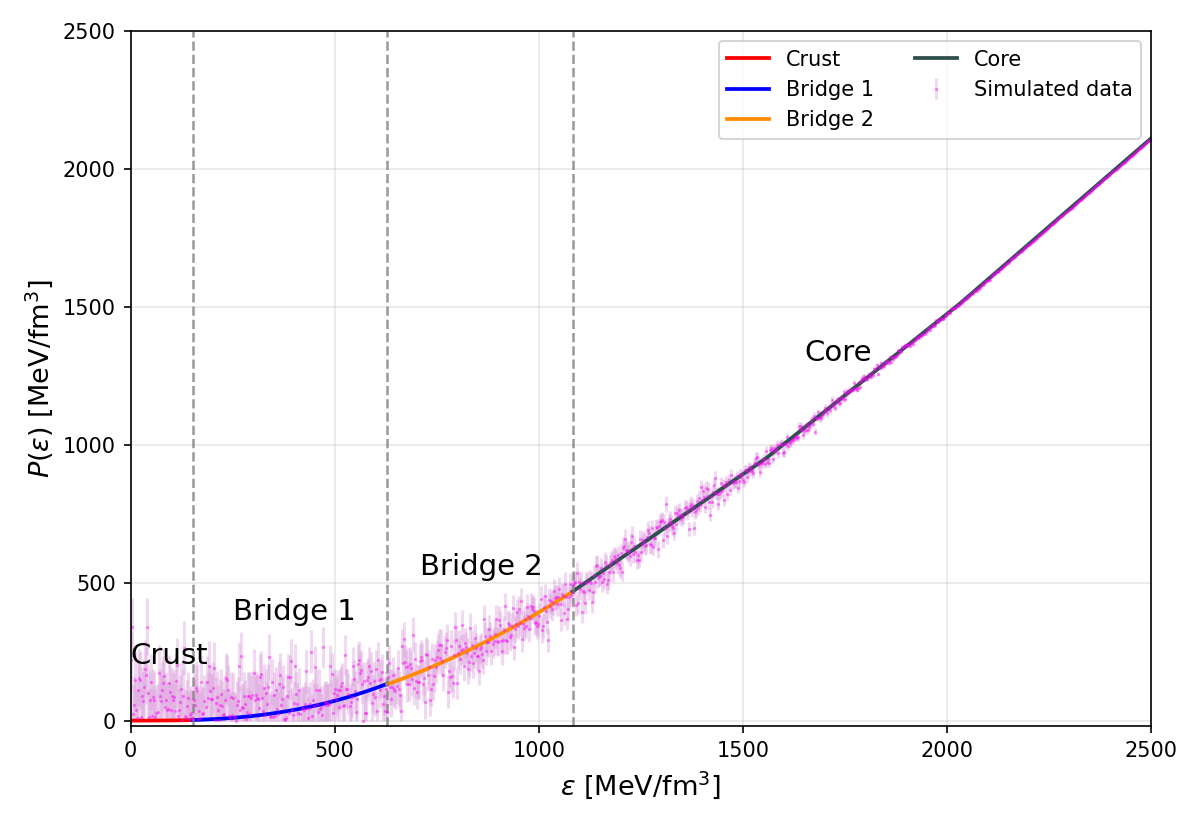}
    \caption{Illustration of the piecewise continuous parametrization of the EoS. The solid curves show the crust, bridge and core segments, while the magenta points represent the simulated EoS data with injected Gaussian noise. The vertical dashed lines indicate the three matching energy densities separating the four regions. Continuity is imposed at each segment boundary, producing a smooth representation of the complete EoS.}
    \label{fig:placeholder}
\end{figure}
\section{Evaluating the Analytical Fitting Forms}
\label{sec:analytical_fit_evaluation}
The fitting procedure seeks to find, for a particular EoS table, the fitting parameter vector $\boldsymbol{\theta_0}$. In the crust region of $0 < \mathcal{E} < \mathcal{E}_1$ one defines a $\chi^2_{\text{crust}}$ measure as
\begin{align}
    \chi^2_{\text{crust}} = \sum_{i = 1}^{N}\bigg[P_i - \mathcal{P}_\text{crust}(\mathcal{E}_i, \theta_{\text{crust}})\bigg]^2
\end{align}
where the summation index $i$ runs over entries of the EoS table itself, $P_i = P(\mathcal{E}_i)$ is the pressure for energy density $\mathcal{E} = \mathcal{E}_i$ sampled from the EoS table and $\mathcal{P}_{\text{crust}}(\mathcal{E}_i, \theta_\text{crust})$ is the pressure obtained from the fitting function for the same value of $\mathcal{E}_i$. By minimizing $\chi^2_\text{crust}$, one finds an initial guess on $\theta_\text{crust} = \{K, \alpha\}$. 


\par For the intermediate bridge regions, guesses on the coefficients $\{a_i, b_i\}$ are obtained by requiring that at the edges of the bridges, the pressure obtained from the fitting function should match the pressure obtained from the EoS table. This is made sure for the $i^{\text{th}}$ bridge by solving the following set of linear algebraic equations 
\begin{align}
    a_i\mathcal{E}^{\alpha_i}_{i} + b_i \mathcal{E}^{\beta_i}_{i} = P(\mathcal{E}_i)\\
    a_i\mathcal{E}^{\alpha_i}_{i+1} + b_i \mathcal{E}^{\beta_i}_{i+1} = P(\mathcal{E}_{i+1})
\end{align}
for $i = 1, 2, 3$ and for properly adjusted values of $\alpha_i$ and $\beta_i$. For $\mathcal{E} > \mathcal{E}_3$ (or $\mathcal{E}_4$ depending on the particular EoS in question), the EoS is parametrized using a piecewise constant speed-of-sound construction. The $\mathcal{E} > \mathcal{E}_{3(4)}$ region is divided into $N_{\mathrm{seg}}$ number of segments defined by the energy-density nodes $\{\mathcal{E}_j\}_{j = 3(4)}^{N_{\text{seg}}}$. Within each segment, the squared speed of sound $c_{s,j}^{2}$ (for the segment $\mathcal{E}_j \leq \mathcal{E} < \mathcal{E}_{j+1}$, $j \geq 3(4)$) is assumed to be constant and approximately determined from calculating the local slope from table as 
\begin{align}
    c^2_{s,j} = \frac{P_{j+1} - P_j}{\mathcal{E}_{j+1} - \mathcal{E}_{j}}
\end{align}
One models the equation for the $j^{\text{th}}$ segment
\begin{equation}
    \mathcal{P}(\mathcal{E}) = P_j + c_{s,j}^{2}
    \left(
    \mathcal{E} - \mathcal{E}_j
    \right),
    \qquad
    \mathcal{E}_j \leq \mathcal{E} < \mathcal{E}_{j+1}.
\end{equation}

The nodal pressures $\{P_j\}$ are determined recursively by enforcing continuity of the pressure across successive nodes. So the EoS beyond $\mathcal{E}_{3}$(or $\mathcal{E}_{4}$) is represented by a sequence of linear segments whose slopes are determined by the values of $c_{s,j}^{2}$. At the end of these numerical calculations, on gets the full parameter vector $\boldsymbol{\theta} = \boldsymbol{\theta_0}$, and hence, the the analytical fitting function $\mathcal{P} = \mathcal{P}(\mathcal{E}, \boldsymbol{\theta_0})$. The parameter vector obtained in this way reproduces the tabulated EoS with sufficient degree of accuracy. 

\section{Synthetic and Multi-Messenger Inference}
The EoS fitting function is obtained from a single deterministic fit to the table, with no associated uncertainties. On the other hand, inferring NS EoSs from MM observations is essentially an inverse problem. From macroscopic observables with various sources of noise, systematic, statistical uncertainties, and finite precision of measurement, one tries to infer properties of the underlying EoS. However, there are two serious hurdles in this approach. Firstly, multiple distinct models can produce nearly identical predictions for NS observables. Moreover, the presence of these uncertainties further implies that multiple EoSs can remain consistent with any given MM dataset within its error bars. One should, therefore, aim to pin down a distribution of EoSs rather than search for a single underlying EoS consistent with MM measurements. This is particularly important given the limited number of currently available MM observations. As more events are detected, the distribution is expected to become progressively narrower.
In this work, we address these uncertainties and the resulting non-uniqueness of the inferred EoS in two complementary ways within a Bayesian framework. In the first approach, we construct synthetic or mock EoSs by deliberately injecting Gaussian noise into the true EoSs, with a finite noise-distribution width to mimic realistic observational uncertainty. Using log-likelihood maximization we can obtain the median parameters, the corresponding posterior distribution and the corresponding EoSs around the true EoS. This analysis serves as a recoverability check of the fitting parameters against noise.

\par The first approach, however, does not reveal how uncertainties in MM measurements affect the inferred posterior distribution of the EoS. In the second approach, we do not impose a specific noise model on the true EoS. Instead, we allow the MM observations, each with its own reported uncertainties and sensitivity to different regions of the EoS parameter space, to directly constrain the posterior distributions of the model parameters. Given the limited number of observed ${\sim}2~M_\odot$ pulsars and the availability of only one GW event with a well-constrained tidal deformability, GW170817, the MM-driven posteriors may be somewhat broader. This comparison therefore, provides a useful means of assessing how observational uncertainties affect constraints on the NS EoS.

\par  Before we discuss in the next sections the method of noise insertion, MM log likelihood maximization and the Bayesian frameworks associated with them, it is important to point out that not all the components of the parameter vector $\boldsymbol{\theta}$ are independent. Due to continuity conditions on the pressure at each matching point, $K$ and $b_2$ can actually be obtained from the other parameters. Therefore, the independent parameters are $\boldsymbol{\theta} - \{K, b_2\}$. Hence, from this point onward, by the symbol $\boldsymbol{\theta}$, we will refer to this reduced set of independent parameters. 

\section{Bayesian Inference Framework}
\label{sec:bayes_framework}

\par The following phase of this work is to use the fitting parameters of a particular tabulated EoS as injection values within a Bayesian inference framework, proceeding in two steps: (i) we inject Gaussian noise into the true EoS to obtain the median and posterior EoS samples, thereby verifying recoverability by comparing the resulting median EoS against the true EoS; and (ii) we construct a Bayesian framework incorporating MM data to obtain the median and posterior EoS samples around the true EoS. We use Bayes' theorem,
\begin{equation}
  p\!\left(\boldsymbol{\theta} \mid \{d_k\}\right)
  \;\propto\;
  p(\boldsymbol{\theta})\;
  \mathcal{L}\!\left(\boldsymbol{\theta} \mid \{d_k\}\right),
  \label{eq:bayes}
\end{equation}
where $p(\boldsymbol{\theta})$ is the prior distribution encoding our belief about the EoS parameters before data are considered, $\mathcal{L}(\boldsymbol{\theta}\mid\{d_k\})=p(\{d_k\}\mid\boldsymbol{\theta})$ is the likelihood function and $p(\boldsymbol{\theta}\mid\{d_k\})$ is the posterior distribution. The posterior is explored using the nested sampling algorithm \textsc{Dynesty}~\citep{Speagle:2020} as implemented in the \textsc{Bilby} inference library~\citep{Ashton:2018}. In sec.\ref{synthetic} and sec.\ref{multimessenger}, we discuss in detail the construction of the likelihood functions for both Syn data and MM data driven Bayesian inference respectively.

\subsection{Synthetic EoS Generation}
\label{synthetic}

\par For a particular tabulated EoS table $\{\mathcal{E}_i, P_i\}_{i = 1}^{N}$, we evaluate the analytical fitting function: $\mathcal{P}_i = \mathcal{P}(\mathcal{E}_i; \boldsymbol{\theta})$ on the grid $\{\mathcal{E}_i\}_{i = 1}^{N}$ of the EoS table. To simulate realistic sources of uncertainties and construct synthetic EoSs, at each $\mathcal{E}_i$ we add Gaussian noise as
\begin{align}
    \overline{\mathcal{P}}_i = \mathcal{P}_i + \mathcal{N}(0, \sigma^2_i)
\end{align}
where $\sigma_i = 0.3\times \text{max}(|\mathcal{P}(\mathcal{E}_{N-i+1})|, 10^{-8})$ with $i=1,\dots,N$ and generate mock EoSs around the tabulated EoS. The floor value $10^{-8}$ is introduced to avoid numerical ill-conditioning at vanishing pressures. We define the log likelihood function as

\begin{align}
    \text{ln}~\mathcal{L}_{\text{Syn}} = -\sum_{i = 1}^{N} \bigg[
    \frac{(\overline{\mathcal{P}}_i - \mathcal{P}_i)^2}{\sigma^2_i} + \frac{1}{2}\text{ln}(2\pi\sigma^2_i)\bigg]
\end{align}
We use $\boldsymbol{\theta_0}$ as the injected model parameter, we choose Gaussian priors with mean as the components of $\boldsymbol{\theta_0}$ with sufficiently broad prior widths. By maximizing $\text{ln}\mathcal{L}_{\text{Syn}}$ yields the full marginalized posterior distribution for each fitting parameter, from which we report the median values and associated credible intervals and posterior distributions, and hence, the median EoS and posterior EoS samples. 

\begin{table*}[t]
\centering
\renewcommand{\arraystretch}{1.0}
\setlength{\tabcolsep}{2pt}
\begin{adjustbox}{max width=\textwidth,center}
\begin{tabular*}{\textwidth}{@{\extracolsep{\fill}}l c c c c@{}}
\toprule
\toprule
\textbf{Stellar Object} & \textbf{$M/M_\odot$} & \textbf{$R$ (km)} & \textbf{$\Lambda$} & \textbf{$I$ ($10^{45} \, \mathrm{g\cdot cm^2}$)} \\
\midrule
\midrule

PSR J0030+0451 & $1.34^{+0.15}_{-0.16}$~\cite{Riley:2019yda,Miller:2019cac} & $12.71^{+1.14}_{-1.19}$~\cite{Riley:2019yda,Miller:2019cac} & $370^{+360}_{-130}$~\cite{Jiang:2019rcw} & $1.43^{+0.30}_{-0.13}$~\cite{Jiang:2019rcw} \\
PSR J0348+0432 & 1.97--2.05~\cite{Antoniadis:2013pzd} & 12.16--12.948~\cite{Huo:2018dqu} & $-$ & 1.594--1.9073~\cite{Zhao:2016rfv} \\
PSR J0740+6620 & $2.08\pm 0.07$~\cite{Miller:2021qha,Fonseca:2021wxt} & $12.39^{+1.30}_{-0.98}$~\cite{Riley:2021pdl} & $575^{+262}_{-232}$~\cite{Biswas:2021yge} & $4.65^{+1.16}_{-0.82}$~\cite{Li:2022ozz} \\
PSR J1614-2230 & $1.97\pm 0.04$~\cite{Ozel:2010bz,Demorest:2010bx} & $8.6-13.8$~\cite{Majid:2020hlg} & $-$ & $-$ \\
PSR J0437--4715 & $1.418\pm 0.037$~\cite{Choudhury:2024xbk} & $11.36^{+0.95}_{-0.63}$~\cite{Choudhury:2024xbk} & $-$ & $-$ \\
PSR J1231-1411 & $1.12\pm 0.07$~\cite{Qi:2025mpn} & $9.91^{+0.88}_{-0.86}$~\cite{Qi:2025mpn} & $-$ & $-$ \\
PSR J0952--0607 A & $2.35\pm 0.17$~\cite{Romani:2022jhd} & $-$ & $-$ & $-$ \\

\multirow[c]{2}{*}{
$\begin{array}{l}{\hspace{-0.2cm}}
\text{PSR J0737-3039} {\hspace{0.5cm}}
\begin{array}{l} 
\text{A} \\
\text{B}
\end{array}
\end{array}$
}
& $1.377\pm 0.005$~\cite{Lyne2006}
& $8.14-25.74$~\cite{Yang}
& $-$
& $1.15^{+0.38}_{-0.24}$~\cite{Landry:2018jyg} \\
& $1.250\pm 0.005$~\cite{Lyne2006}
& $-$
& $-$
& $-$ \\
\multirow[c]{2}{*}{
$\begin{array}{l}{\hspace{-0.2cm}}
\text{GW170817} 
{\hspace{0.5cm}}
\begin{array}{l}
\text{Primary} \\
\text{Secondary}
\end{array}
\end{array}$
}
& $1.46^{+0.12}_{-0.10}$~\cite{LIGOScientific:2017vwq}
& {$10.8^{+2.0}_{-1.7}$~\cite{LIGOScientific:2018cki}}
& \multirow[c]{2}{*}{$\tilde{\Lambda}~\approx~  300^{+420}_{-230}$~\cite{LIGOScientific:2018hze}}
& \multirow[c]{2}{*}{$1.43^{+0.30}_{-0.13}$~\cite{Jiang:2019rcw}} \\
& $1.27^{+0.09}_{-0.09}$~\cite{LIGOScientific:2017vwq}
& {$10.7^{+2.1}_{-1.5}$~\cite{LIGOScientific:2018cki}} & & \\

\multirow[c]{2}{*}{
$\begin{array}{l}{\hspace{-0.2cm}}
\text{GW190425} 
{\hspace{0.5cm}}
\begin{array}{l}
\text{Primary} \\
\text{Secondary}
\end{array}
\end{array}$
}
& $2.1^{+0.5}_{-0.4}$~\cite{LIGOScientific:2020aai}
& \multirow[c]{2}{*}{$-$}
& \multirow[c]{2}{*}{$1.4^{+3.8}_{-1.2}\times 10^3$~\cite{Han:2020qmn}}
& \multirow[c]{2}{*}{$-$} \\

& $1.3^{+0.3}_{-0.2}$~\cite{LIGOScientific:2020aai}
& & & \\

\bottomrule
\bottomrule
\end{tabular*}
\end{adjustbox}
\caption{Summary of masses ($M/M_\odot$), radii ($R$), tidal deformabilities ($\Lambda$), and inferred moments of inertia ($I$) of various pulsars, GW events and other compact objects. The $M$-$R$ constraints for 4U 1724-207, 4U 1608-52 and EXO 1745-248 are taken from Ref.~\cite{Ozel:2015fia}. Similar constraints for HESS J1731-347 and PSR J1824-2452I (M28I) are taken from Refs.~\cite{Doroshenko:2022nwp} and~\cite{Vurgun:2022vvo} respectively. In our comparison with theoretical predictions, we have projected the moment of inertia (MoI) of PSR J0030+0451~\cite{Jiang:2019rcw}, estimated at a mass of $M = 1.4~M_\odot$, across the full range of mass and radius estimates~\cite{Riley:2019yda,Miller:2019cac}.}
\label{tab:Data}
\end{table*}

\subsection{Multi-Messenger Driven inference}
\label{multimessenger}


We consider a set of NS observations, each labeled by an index $k$, as shown in Table.~\ref{tab:Data}, each providing a measurement of stellar mass $M_{\rm obs}^{(k)}$ and, for a subset of sources, a measurement of tidal deformability $\Lambda_{\rm obs}^{(k)}$ at the benchmark mass of $M = 1.4~M_\odot$. Almost all observational uncertainties are asymmetric; the lower and upper $1\sigma$ error bars are denoted $\sigma_{-}^{(k)}$ and $\sigma_{+}^{(k)}$, respectively, for the $k^{\text{th}}$ observation.

\subsubsection{Construction of Likelihood Function}
Now for each observed NS $k$, the corresponding central energy density $\mathcal{E}_c$ is completely unknown. Given an EoS, the central density uniquely determines all stellar properties. However, because $\mathcal{E}_c$ is not directly observable, it must be treated as a  nuisance parameter and should be marginalized over. The marginalized likelihood for $k^{\text{th}}$ star is
\begin{equation}
\begin{aligned}
\mathcal{L}\!\left(d_k \mid \boldsymbol{\theta}\right)
&= \int
   \mathcal{L}\!\left(d_k \mid \boldsymbol{\theta},\,\mathcal{E}_c\right)\,
   p\!\left(\mathcal{E}_c\right)\,
   d\mathcal{E}_c \\
&= \int
   \mathcal{L}\!\left(d_k \mid \boldsymbol{\theta},\,M\right)\,
   p\!\left(M\right)\,
   \bigg|\frac{dM}{d\mathcal{E}_c}\bigg|d\mathcal{E}_c ,
\end{aligned}
\label{eq:marg_integral}
\end{equation}
where $\mathcal{L}(d_k\mid\boldsymbol{\theta},\mathcal{E}_c)$ is the likelihood of the data given the EoS and the specific central density,
and $p(\mathcal{E}_c)$ is the prior on central density. In going from the first to the second line, we have used the jacobian transformation of the prior
\begin{align}
    p\!\left(\mathcal{E}_c\right)\, d\mathcal{E}_c = p\!\left(M\right)\bigg|\frac{dM}{d\mathcal{E}_c}\bigg|d\mathcal{E}_c
\end{align}
Now in the discrete limit the above equation becomes
\begin{align}
    p\!\left(\mathcal{E}_{c,i}\right)\, \Delta\mathcal{E}_{c,i} = p\!\left(M\right)\bigg|\frac{dM}{d\mathcal{E}_c}\bigg|_{i}\Delta\mathcal{E}_{c,i} = p\!\left(M\right) w_i
\end{align}
The weight $w_i = \big|\frac{dM}{d\mathcal{E}_c}\big|_{i}\Delta\mathcal{E}_{c,i}$ encodes the prior probability assigned to central density $\Delta\mathcal{E}_{c,i}$. The Jacobian $|dM/d\Delta\mathcal{E}_{c,i}|_i$ is computed numerically via a central finite-difference scheme applied to the TOV. A naive uniform prior on $\mathcal{E}_{c}$ assigns equal weight $w_i\propto\Delta\mathcal{E}_{c,i}$ to all grid points. Near the maximum-mass end of the stable branch, the mapping $\varepsilon_c\to M$ becomes very flat, so a uniform prior on $\varepsilon_c$ over-represents high-density configurations that all correspond to nearly the same mass. We instead adopt a uniform prior on mass, which is the least informative choice in the absence of population data. Then setting $P(M) = \text{constant}$, we get the likelihood function in the discrete limit as
\begin{equation}
  \mathcal{L}\!\left(d_k \mid \boldsymbol{\theta}\right)
  = \frac{1}{W}
    \sum_{i\in\mathcal{S}}
    p\!\left(M_{\rm obs}^{(k)} \mid M_i\right)\;
    p\!\left(\Lambda_{\rm obs}^{(k)} \mid \Lambda\right)w_i,
  \label{eq:marg_discrete}
\end{equation}
where $W = \sum_{t\in S}w_i$ is used to correctly normalize the likelihood, the sum runs over all stable-branch grid points $i$ in the set
\begin{equation}
  \mathcal{S}
  = \bigl\{
      i \;:\;  M_{\rm min} \leq M_i \leq M_{\rm max, eos} \bigr\},
  \label{eq:S_set}
\end{equation}
and $M_i\equiv M(\varepsilon_{c,i})$ and $\Lambda$, the theoretical prediction of tidal deformability at $M = 1.4M_\odot$, are the EoS-predicted mass at grid point $i$ and tidal deformability, obtained directly from the TOV integration without any approximation.
%
This weight $w_i$ automatically suppresses the high-density tail near $M_{\rm max}$ where $dM/d\varepsilon_c\to 0$, and the factor $W$ in the denominator of Eq.~\eqref{eq:marg_discrete} ensures the result is invariant to the choice of grid resolution $N_{\rm eos}$.
\par The integration domain spans the valid stable branch of the star. However, for ease of numerical analysis and to reduce computational cost, in our numerical analysis we have chosen that range of $\mathcal{E}_c$ which corresponds to $M\in[M_{\rm min} = 1.25~M_\odot,\,M_{\rm max}]$, $\,M_{\rm max}$ being the maximum mass of the star obtained from the EoS. 
\subsubsection{Asymmetric Measurement Uncertainties: Split-Normal Likelihood}

Reported uncertainties on NS masses and tidal deformabilities are generally asymmetric, reflecting skewed posterior distributions. To preserve this asymmetry, we adopt the split-normal distribution as our single-observable likelihood,
\begin{equation}
  \mathrm{SN}(x \mid x_{\rm obs},\,\sigma_{-},\,\sigma_{+})
  = \frac{\sqrt{2/\pi}}{\sigma_{-}+\sigma_{+}}
    \exp\!\left[-\frac{(x-x_{\rm obs})^{2}}{2\,\sigma(x)^{2}}\right],
  \label{eq:split_normal}
\end{equation}
where the local standard deviation is
\begin{equation}
  \sigma(x)
  = \begin{cases}
      \sigma_{-} & x < x_{\rm obs}, \\[4pt]
      \sigma_{+} & x \geq x_{\rm obs}.
    \end{cases}
  \label{eq:split_sigma}
\end{equation}
%
The per-point joint mass log-likelihood at stable-branch grid point $i$ for observation $k$ is therefore
%

\begin{align}
    \ell_{M,i}^{(k)}=
\log \mathrm{SN}\!\left(
M_i \mid M_{\rm obs}^{(k)},
\sigma_{M,-}^{(k)},\,
\sigma_{M,+}^{(k)}
\right)
\end{align}
Similar log-likelihood for $\Lambda_{1.4}$ is given by 
\begin{align}
    \ell_{\Lambda}^{(k)}=
\log \mathrm{SN}\!\left(
\Lambda \mid \Lambda_{\rm obs}^{(k)},
\sigma_{\Lambda,-}^{(k)},\,
\sigma_{\Lambda,+}^{(k)}
\right)
\end{align}
if for the $k^{\text{th}}$ observation, the measurement of $\Lambda_{1.4}$ is available. $\Lambda$ is the dimensionless tidal deformability calculated as $M = 1.4M_\odot$ benchmark mass.
\subsubsection{Total Log-Likelihood}
Assuming all the NS observations are statistically independent -they are drawn from different astrophysical systems measured by different instruments -the total likelihood factorises as a product over stars,
\begin{equation}
  \mathcal{L}\!\left(\boldsymbol{\theta}\mid\{d_k\}\right)
  = \prod_{k=1}^{K} \mathcal{L}\!\left(d_k\mid\boldsymbol{\theta}\right).
  \label{eq:likelihood_product}
\end{equation}
Taking logarithms and substituting the discrete marginalization of Eq.~\eqref{eq:marg_discrete},
\begin{equation}
\begin{aligned}
\log\mathcal{L}_{\rm total}
=\sum_{k=1}^{N}
\Bigg[
\log\sum_{i\in\mathcal{S}}{}&
\exp\!\Big(
    \ell_{M,i}^{(k)}
    +\log w_i
\Big)
\\
&\qquad
+\delta^{(k)}\ell_{\Lambda}^{(k)}
-\log W
\Bigg].
\end{aligned}
\label{eq:logL_total}
\end{equation}

%
where $\delta^{(k)}$ is defined as
\begin{align}
    \delta^{(k)} &= 1 \qquad \text{if the measurement of $\Lambda_{1.4}$ is available}\\ \nonumber
    &=0 \qquad \text{otherwise} \nonumber
\end{align}
The following hard cuts are imposed as part of the likelihood evaluation
\begin{enumerate}
  \item The matching densities must be strictly ordered:
        $\mathcal{E}_{1}
         <\mathcal{E}_{2}
         <\mathcal{E}_{3}
         (<\mathcal{E}_{4})
         <\mathcal{E}_{\text{max}}$.
  \item For a particular EoS, only those measurements will be considered in the Bayesian analysis for which $M^{(k)} \leq M_{\text{max},\text{EoS}}$, $M^{(k)} \leq M_{\text{max},\text{EoS}}$ being the the maximum stellar mass obtain from the EoS.

  \item The pressure must be non-negative everywhere on the EoS grid: $P(\mathcal{E}_i)\geq 0$ $\forall$ $i$.
  \item The pressure must be monotonically non-decreasing: $dP/d\mathcal{E}\geq 0$, ensuring thermodynamic stability.
  \item Pressure continuity must hold at all the matching points.
  \item The TOV integration must converge with at least five valid central-density solutions.
  \item The stable branch, after applying the cuts $M\geq M_{\rm min} = 1.25~M_\odot$ and $\Lambda\geq\Lambda_{\rm min} = 2$ (must be non-empty).
\end{enumerate}

\section{Results}
\label{sec:results}
\begin{figure}[t]
    \centering
    \includegraphics[width=0.48\textwidth]{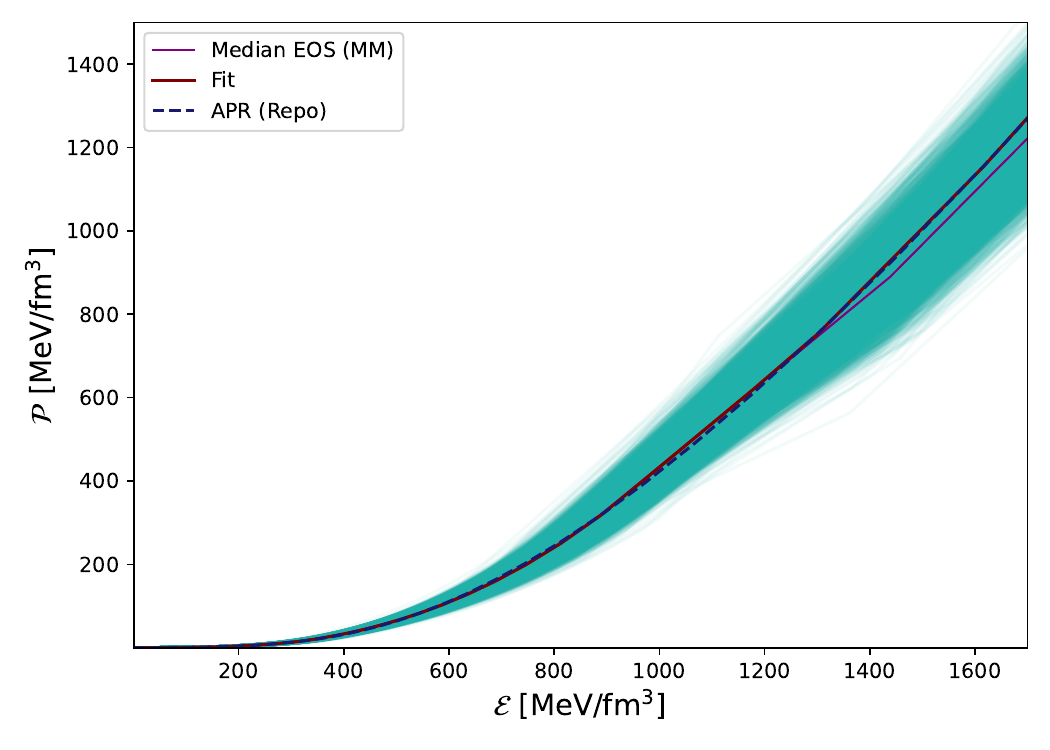}
    \includegraphics[width=0.48\textwidth]{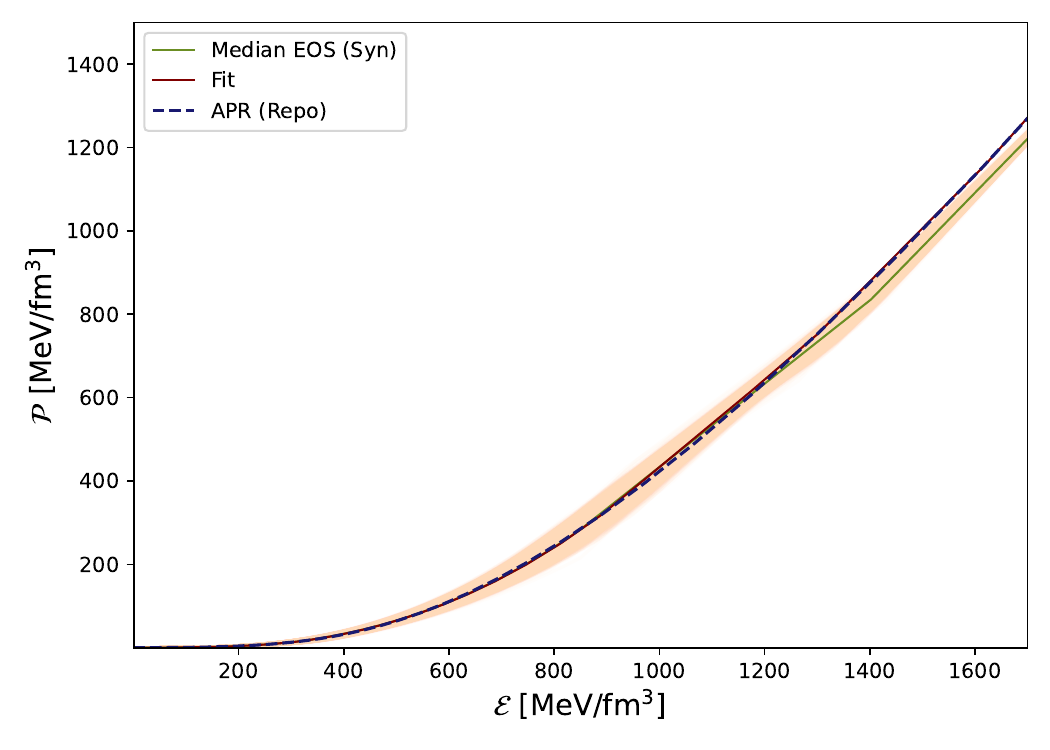}
    \caption{Comparison of the APR with the median and posterior EoSs obtained from likelihood maximization using MM astronomy (top panel) and from synthetic data (bottom panel). The posterior Eos samples from synthetic data generation and MM data driven log-likelihood maximization are also shown.}
    \label{fig:EoS_APR}
\end{figure}

\begin{figure}[t]
    \centering
    \includegraphics[width=0.48\textwidth]{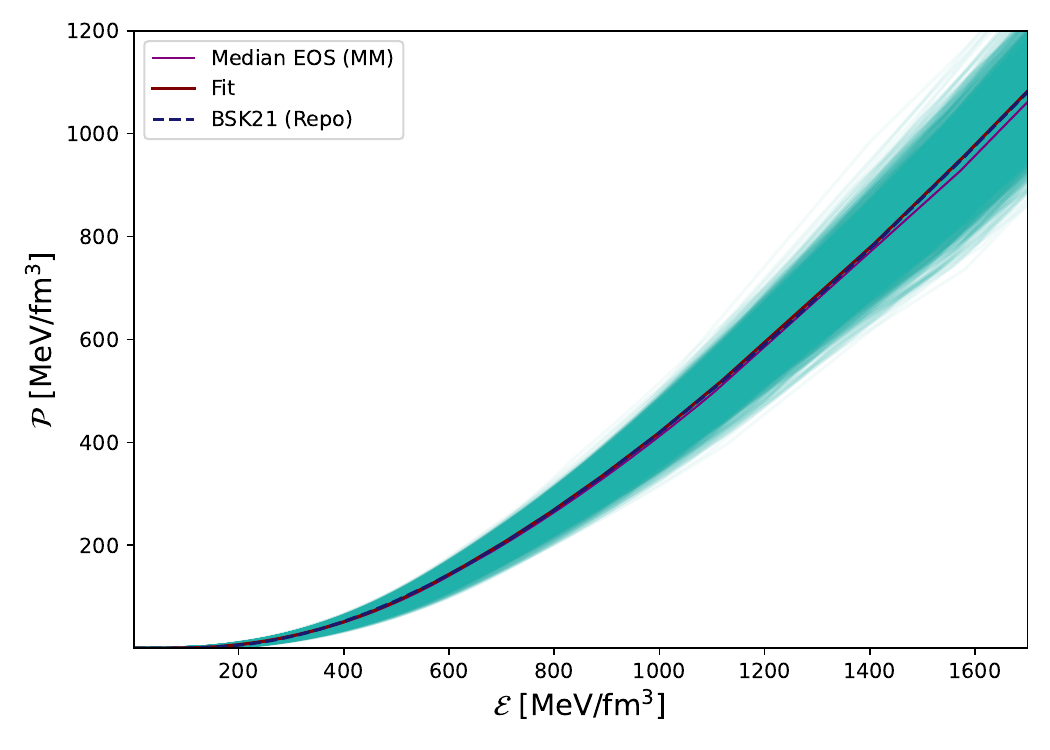}
    \includegraphics[width=0.48\textwidth]{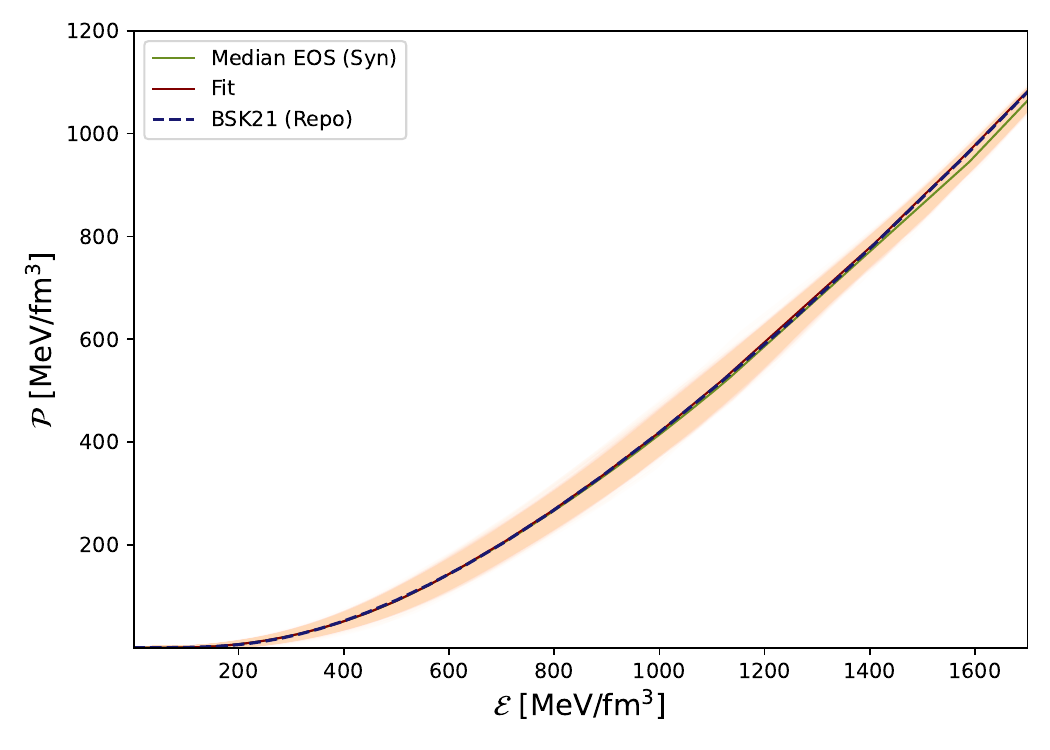}
    \caption{Comparison of the BSK21 with the median and posterior EoSs obtained from likelihood maximization using MM astronomy (top panel) and from synthetic data (bottom panel). The posterior Eos samples from synthetic data generation and MM data driven log-likelihood maximization are also shown.}
    \label{fig:EoS_BSK21}
\end{figure}

\begin{figure*}[t]
    \centering
    \includegraphics[width=0.48\linewidth]{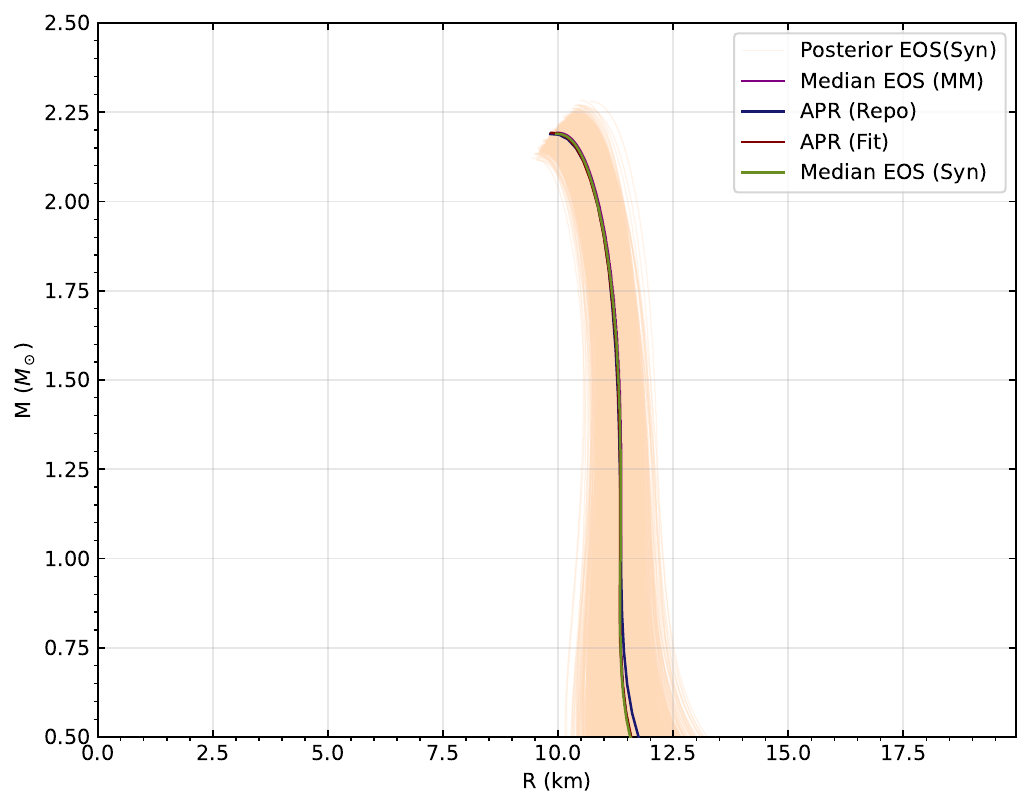}
    \includegraphics[width=0.48\linewidth]{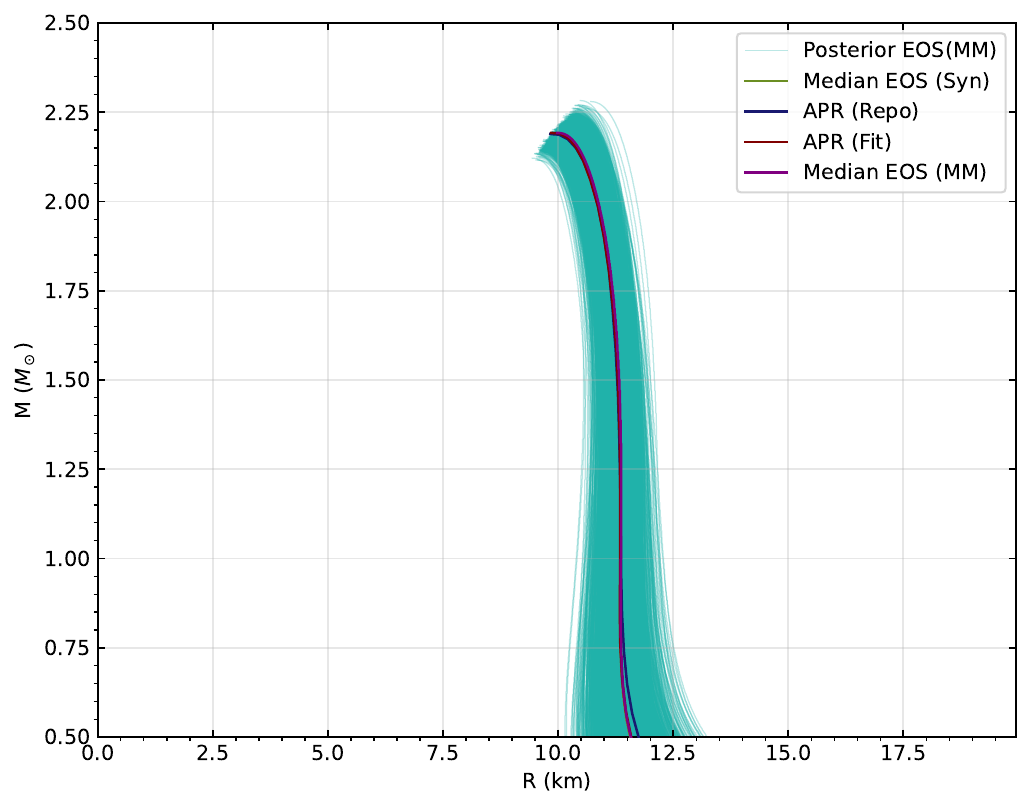}
    \caption{Comparison of the $M$-$R$ relations for the APR EoS. The EoSs are obtained directly from the LALSimulation repository (Repo), from the corresponding fitting functions (Fit), from the median EoS of the Syn data, and from the median EoS obtained through MM data. The results show extremely good agreement among the different representations for both EoSs. The plots also show the posterior MR distributions obtained from Syn data and MM data.}
    \label{fig:MR_comparison_APR}
\end{figure*}

\begin{figure*}[t]
    \centering
    \includegraphics[width=0.48\linewidth]{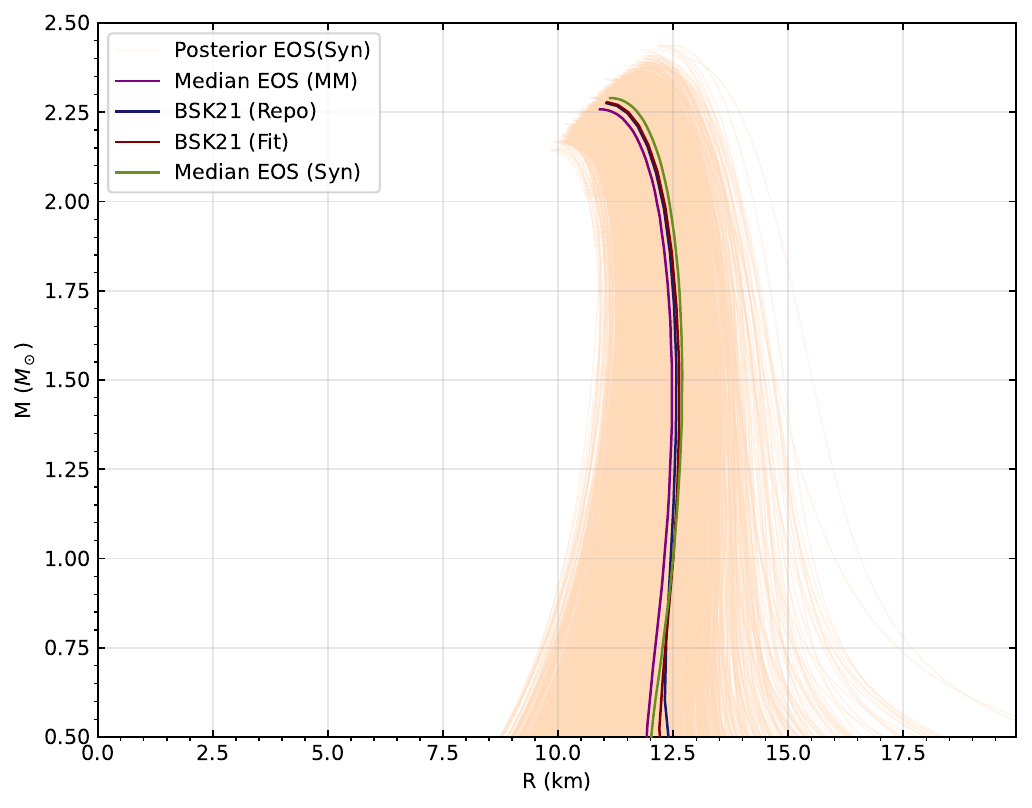}
    \includegraphics[width=0.48\linewidth]{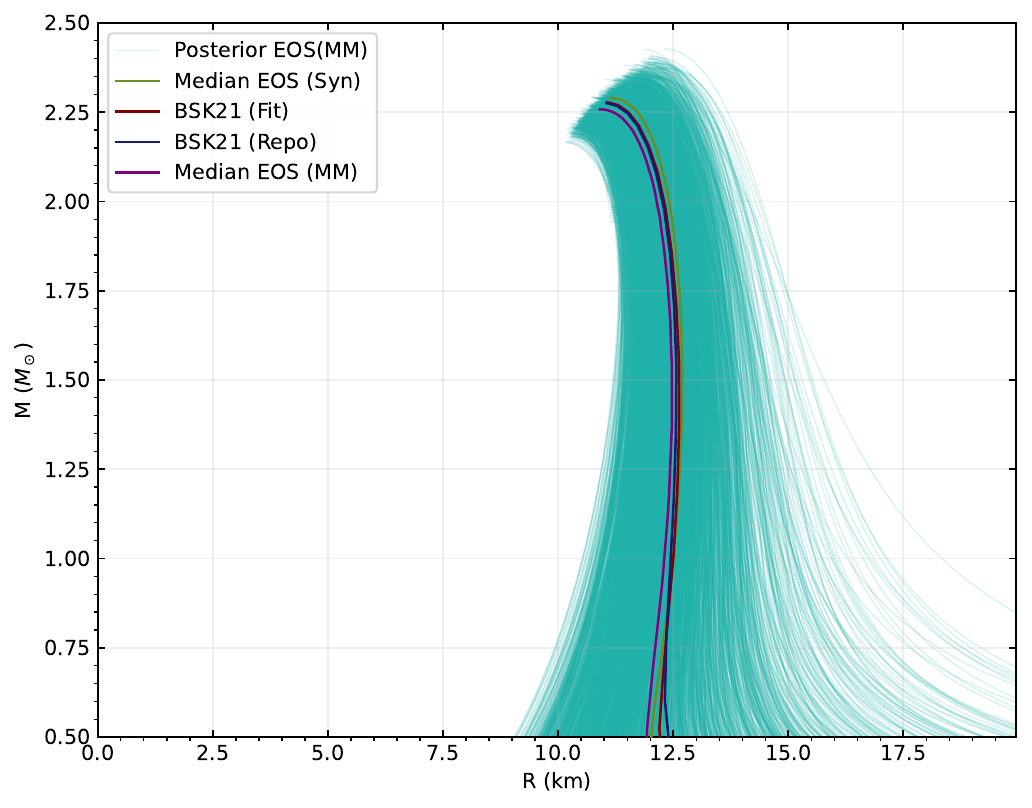}
    \caption{Comparison of the $M$-$R$ relations for the BSK21 EoS. The EoSs are obtained directly from the LALSimulation repository (Repo), from the corresponding fitting functions (Fit), from the median EoS from Syn data, and from the median EoS obtained through MM data. The results show fairly good agreement among the different representations for both EoSs. The plots also show the posterior MR distributions obtained from Syn data and MM data.}
    \label{fig:MR_comparison_BSK21}
\end{figure*}

We first quantify how well the analytical fits reproduce the tabulated EoSs and corresponding macroscopic observables. To show that the present scheme of fitting is not preferential to any particular family of EoS, we illustrate our results through two different families: AP and BSK family.
Fig.~\ref{fig:EoS_APR} and \ref{fig:EoS_BSK21} compare, for the representative cases of APR and BSK21, respectively, the tabulated EoSs obtained from the LALSimulation repository with the EoSs obtained from the analytical fitting forms and the median EoS (i.e. the EoS obtained from the median values of the fitting parameters) obtained from log-likelihood maximization involving MM constraints. We also show compare them with the corresponding posterior EoS samples (i.e. the EoSs obtained from sampling the fitting parameters from their posterior distributions) obtained Syn data generation (peach colored curves) and MM data driven log likelihood maximization (teal colored curves). In the evaluation of the of the posterior samples in both synthetic data driven and MM data driven Bayesian inference, we use the fitting parameter values obtained from the fitting procedure are the injection values and use Gaussian priors for the fitting parameters with injected values as the mean. The corresponding $M$-$R$ relations are shown in Fig.~\ref{fig:MR_comparison_APR} and Fig.~\ref{fig:MR_comparison_BSK21}, for APR and BSK21, respectively, with the Syn data and MM data driven posterior samples plotted using the same colours.

\begin{figure*}[hbt!]
    \centering
    \includegraphics[width=\linewidth]{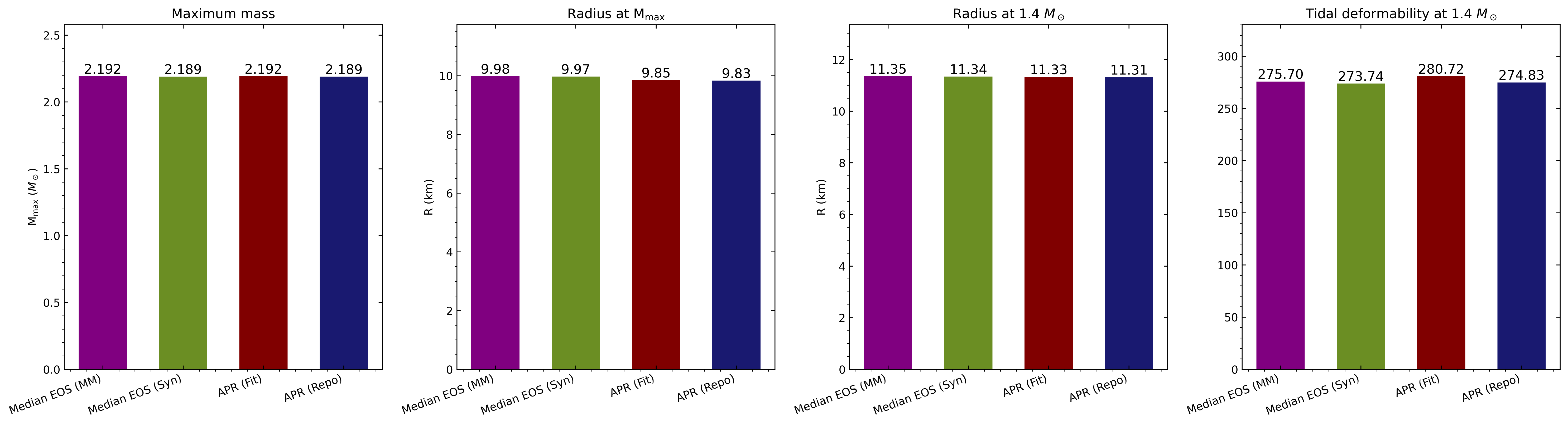}
    \includegraphics[width=\linewidth]{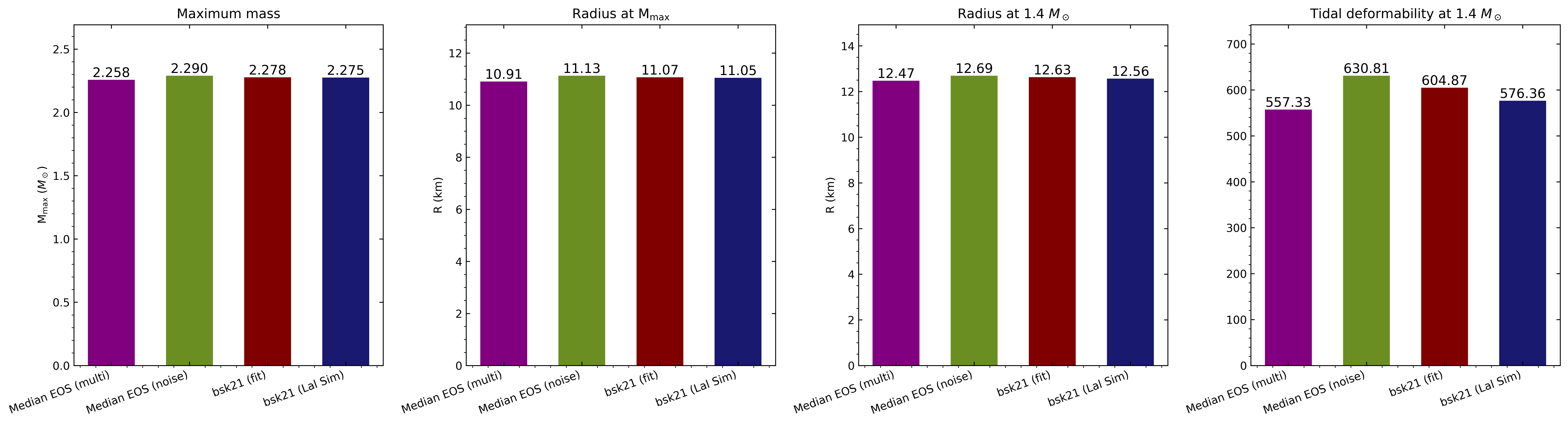}
    \caption{Comparison of $M_{\text{max}}$, stellar radius at $M_{\text{max}}$, radius at $M = 1.4M_\odot$ and $\Lambda_{1.4}$ from the APR EoS (top panel) and from the BSK21 EoS (bottom panel), the analytical fitting form, the median EoS from Syn data and the median EoS obtained MM data.}
    \label{fig:MaxM_Comp}
\end{figure*}

In both cases, the goodness of fit is evident from the figures. The $M$-$R$ relations for APR show remarkable consistency among themselves. The median $M$–$R$ curves obtained from the two phases of log-likelihood analysis lie on top of the $M$-$R$ curves obtained from tabulated APR EoS from LALSimulation repository. The results for BSK21 show a similar level of agreement. For APR, the maximum masses are predicted as $M^{\text{APR}} = 2.189~M_\odot$ obtained from the EoS table from the LALSimulation repository, $M^{\text{APR}}_{\text{Fit}} = 2.192~M_\odot$ obtained from the initially guessed fitting function of the EoS and $M^{\text{APR}}_{\text{median,Syn}} = 2.189~M_\odot$ obtained from the median EoS of the Synj data and $M^{\text{APR}}_{\text{median,MM}} = 2.192~M_\odot$ obtained from the median EoS obtained from MM data, showing excellent consistency among the results.
The MM median result deviates the most from $M^{\text{APR}}$ by $\sim 0.14\%$ only. The stellar radius corresponding to the maximum mass for these four cases are given by $R^{\text{APR}} = 9.83~\text{km}$, $R^{\text{APR}}_{\text{Fit}} = 9.85~\text{km}$, $R^{\text{APR}}_{\text{median,Syn}} = 9.96~\text{km}$ and $R^{\text{APR}}_{\text{median,MM}} = 9.98~\text{km}$ respectively, the MM result deviating the most from $R^{\text{APR}}$ by $\sim 1.5\%$. The dimensionless tidal deformability at $M = 1.4~M_\odot$ benchmark mass for these EoS are given by $\Lambda^{\text{APR}}_{1.4} = 274.83$, $\Lambda^{\text{APR}}_{1.4, \text{median, fit}} = 280.72$, $\Lambda^{\text{APR}}_{1.4, \text{median, Syn}} = 273.74$ and $\Lambda^{\text{APR}}_{1.4, \text{median, MM}} = 275.70$. The stellar radii at $M = 1.4~M_\odot$ are given by $R_{1.4} = 11.31~\text{km}$, ${R_{1.4, \text{fit}}} = 11.33~\text{km}$, $R_{1.4,\text{Syn}} = 11.34~\text{km}$ and $R_{1.4, \text{MM}} = 11.35~\text{km}$. The MM estimate deviates the largest from the actual value of $R_{1.4}$ by $0.35\%$. The values $\Lambda_{1.4}$ obtained from the initially fitted EoS deviate the most from $\Lambda^{\text{APR}}_{1.4}$ by $\sim 2.14\%$, the median values from synthetic EoS generation and MM log likelihood maximization deviating by $\sim 0.4\%$ and $~0.32\%$ respectively. The scale of percentage error for $\Lambda_{1.4}$ is slightly larger than $M_{\text{max}}$ and corresponding radius. This is expected given that $\Lambda\sim C^{-5}$ and slight changes in stellar compactness can shift $\Lambda$ from its actual value. Fig.~\ref{fig:MaxM_Comp} compares $M_{\text{max}}$, the corresponding stellar radius, stellar radius at $M = 1.4~M_\odot$ and $\Lambda_{1.4}$ for APR EoS for all the four EoSs.

\par
\begin{figure*}[t]
    \centering
    \includegraphics[width=0.48\linewidth]{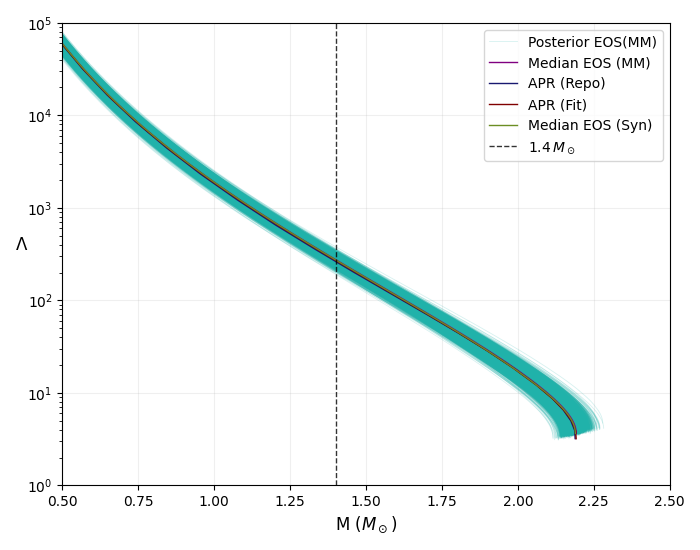}
    \includegraphics[width=0.48\linewidth]{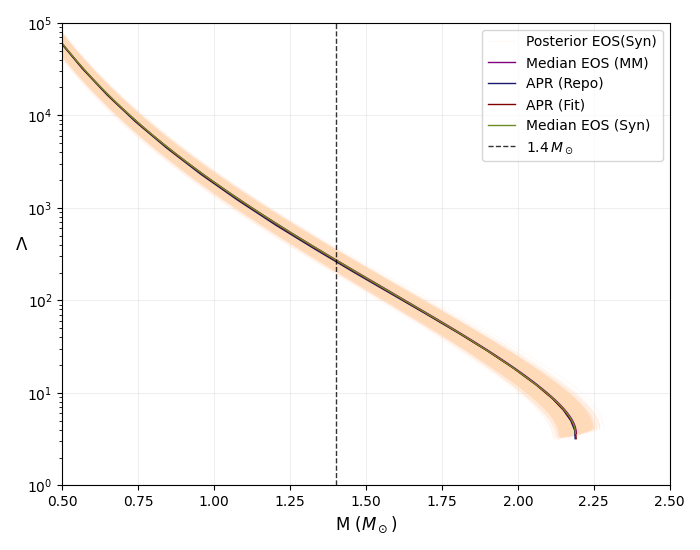}
    \caption{Comparison of the $M$-$\Lambda$ relations for the APR EoS. The results show stellar agreement among the different representations for the same EoSs. The plots also show the posterior $M$-$\Lambda$ distributions obtained from Syn data and MM data.}
    \label{fig:M_Lambda_APR}
\end{figure*}

\begin{figure*}[t]
    \centering
    \includegraphics[width=0.48\linewidth]{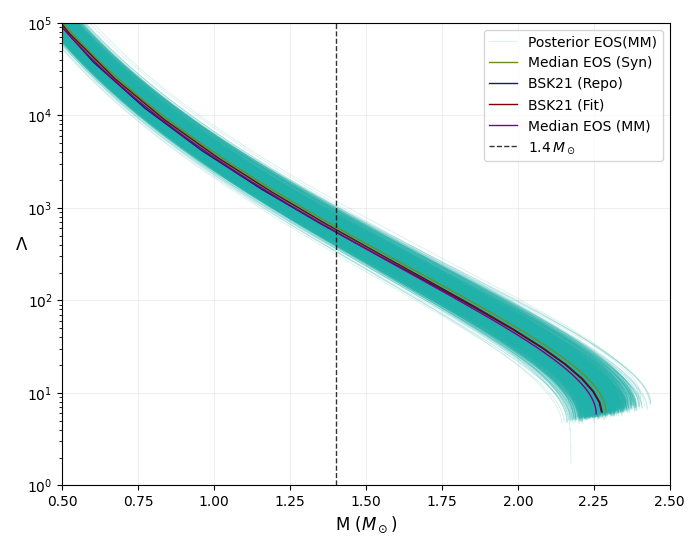}
    \includegraphics[width=0.48\linewidth]{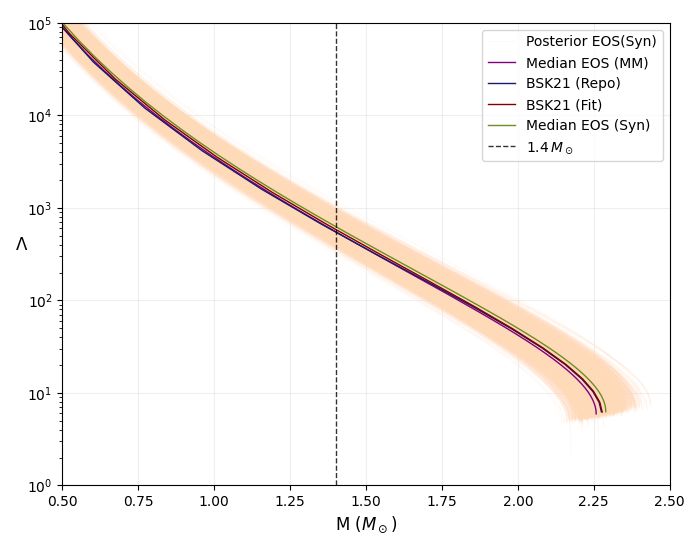}
    \caption{Comparison of the $M$-$\Lambda$ relations for the BSK21 EoS. The results show fairly good agreement among the different representations for the same EoSs. The plots also show the posterior $M$-$\Lambda$ distributions obtained from Syn data and MM data.}
    \label{fig:M_Lambda_BSK21}
\end{figure*}

 Similarly, for BSK21, the maximum masses are evaluated as $M^{\text{BSK21}} = 2.275~M_\odot$ obtained from the EoS table from the LALSimulation repository, $M^{\text{BSK21}}_{\text{Fit}} = 2.278 ~M_\odot$ obtained from the initially guessed fitting function of the EoS and $M^{\text{BSK21}}_{\text{median,Syn}} = 2.290~M_\odot$ obtained from the median EoS of the synthetic EoSs obtained using Syn data and $M^{\text{BSK21}}_{\text{median,MM}} = 2.258~M_\odot$ obtained from the median EoS obtained from MM data, again showing fairly good consistency with the actual BSK21 EoS. Similar to APR, the MM median result deviates the most from $M^{\text{BSK21}}$ by $\sim 0.75\%$ only.  The stellar radius corresponding to the maximum mass for these four cases are given by $R^{\text{BSK21}} = 11.05~\text{km}$, $R^{\text{BSK21}}_{\text{Fit}} = 11.07~\text{km}$, $R^{\text{BSK21}}_{\text{median,Syn}} = 11.13~\text{km}$ and $R^{\text{BSK21}}_{\text{median,MM}} = 10.91~\text{km}$ respectively, the MM result deviating the most from $R^{\text{BSK21}}$ by $\sim 1.27\%$. The stellar radii at $M = 1.4~M_\odot$ are given by $R_{1.4} = 12.56~\text{km}$, ${R_{1.4, \text{fit}}} = 12.63~\text{km}4$, $R_{1.4,\text{Syn}} = 12.69~\text{km}$ and $R_{1.4, \text{MM}} = 12.47~\text{km}$. The prediction of $R_{1.4}$ obtained from the median EoS of the synthetic EoS deviates the most from $R_{1.4}$ of BSK21 by $1.04\%$. The dimensionless tidal deformability at $M = 1.4~M_\odot$ benchmark mass for these EoS are given by $\Lambda^{\text{BSK21}}_{1.4} = 576.36$, $\Lambda^{\text{BSK21}}_{1.4, \text{median, fit}} = 604.87$, $\Lambda^{\text{BSK21}}_{1.4, \text{median, Syn}} = 630.814$ and $\Lambda^{\text{BSK21}}_{1.4, \text{median, MM}} = 557.33$. The values $\Lambda_{1.4}$ obtained from the Syn data deviate the most from $\Lambda^{\text{BSK21}}_{1.4}$ by $\sim 9.45\%$, whereas $\Lambda_{1.4}$ obtained from MM log likelihood maximization performs much better (deviating from the actual value of $\Lambda_{1.4}$ by $3.3\%$). Fig.~\ref{fig:MaxM_Comp} compares $M_{\text{max}}$, the corresponding stellar radius, stellar radius at $M = 1.4~M_\odot$ and $\Lambda_{1.4}$ for BSK21 EoS for all the four EoSs. Table.~\ref{tab:macro-comparison} shows these benchmark data for various families of EoSs. Although we have used APR and BSK21 as benchmark examples for the goodness of fit, median EoSs for all the EoS for which the benchmark data is presented in Table.~\ref{tab:macro-comparison} show similar levels of consistency with the actual EoSs. These comparisons confirm that as far as consistency with macroscopic observables e.g. stellar masses, radii and $\Lambda$ is concerned, the present parametrization is flexible enough to represent both soft EoSs, e.g. the ALF family and very stiff EoSs, e.g. the AP and BSK family, without any bias towards any particular family. Table.~\ref{dataset_table} presents fitting parameters obtained for the fitting function and compares them with the median parameters obtained from synthetic EoS generation and log-likelihood maximization using MM data for various families of Eos. Figs.~\ref{fig:M_Lambda_APR} and \ref{fig:M_Lambda_BSK21} show the $M$-$\Lambda$ relations with the corresponding posterior distributions for APR and BSK21, respectively. The vertical black dotted line denotes $M = 1.4M_\odot$. However, the adopted functional form did not reproduce a few EoSs (GS1, GS2, H1, H2, H3) with sufficient accuracy. A more detailed investigation, potentially employing an alternative functional form, may be required to accurately represent these EoSs.
\section{Parameter Correlations}
\label{sec:correlation}
\begin{figure}[htp!]
    \centering
    \includegraphics[width=\textwidth]{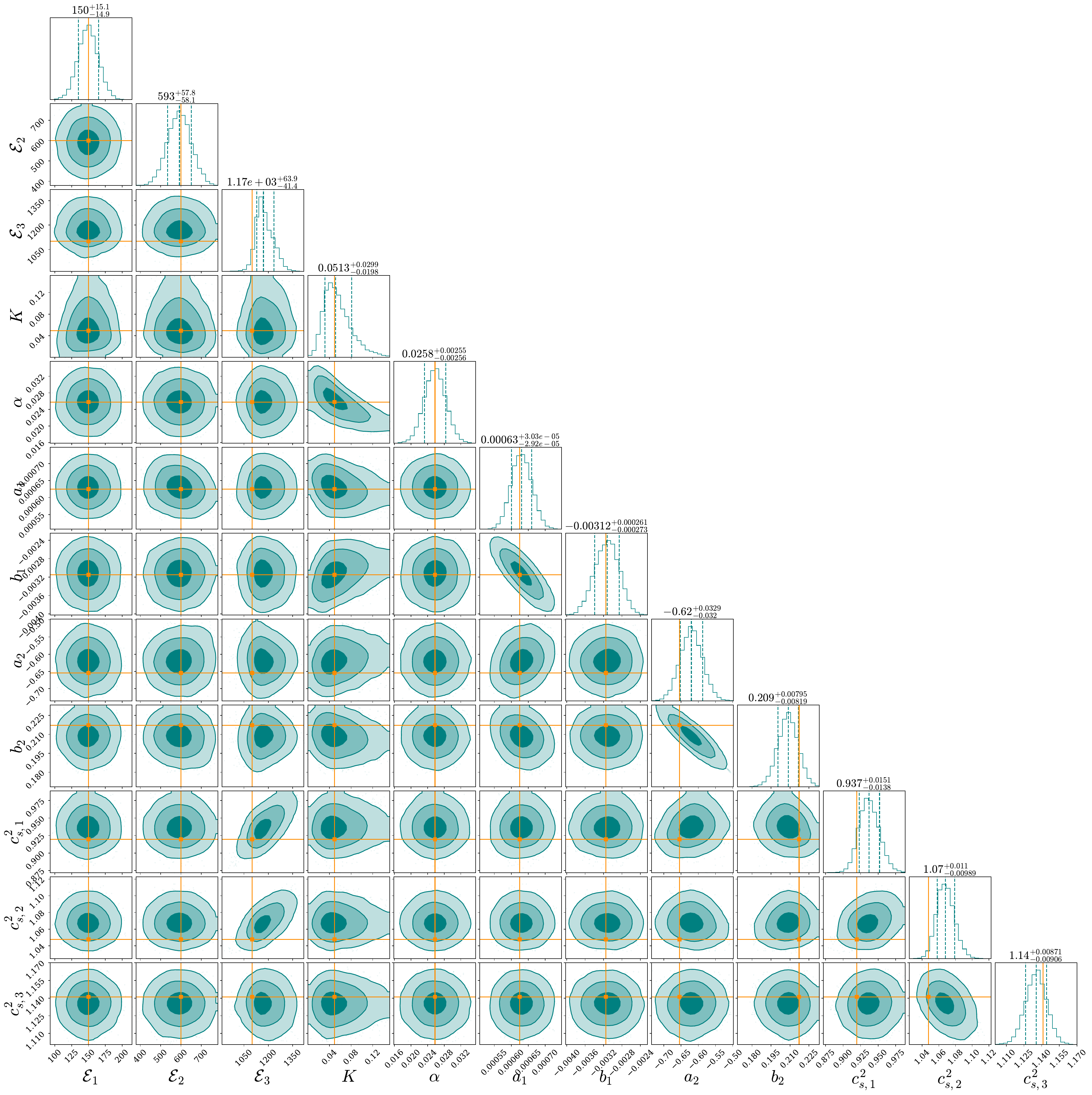}
    \caption{Posterior distributions of the fitting parameters for BSK21 EoS obtained from synthetic EoS generation.}
    \label{fig:BSK21_Fit_Posterior}
\end{figure}

Fig.~\ref{fig:BSK21_Fit_Posterior} presents the posterior distribution of the fitting parameters for BSK21 obtained from the synthetic-data driven likelihood maximization procedure using the fitting parameters as the injection values. During the sampling procedure, we have used Gaussian priors for all the fitting parameters with the injection values of the parameters as the mean. Since in our entire analysis, the bridge exponents $\{ \alpha_i, \beta_i\}$ are fixed, the posterior distributions carry no uncertainty contribution from the exponents.
\par
\begin{figure}[htp!]
    \centering    \includegraphics[width=\textwidth]{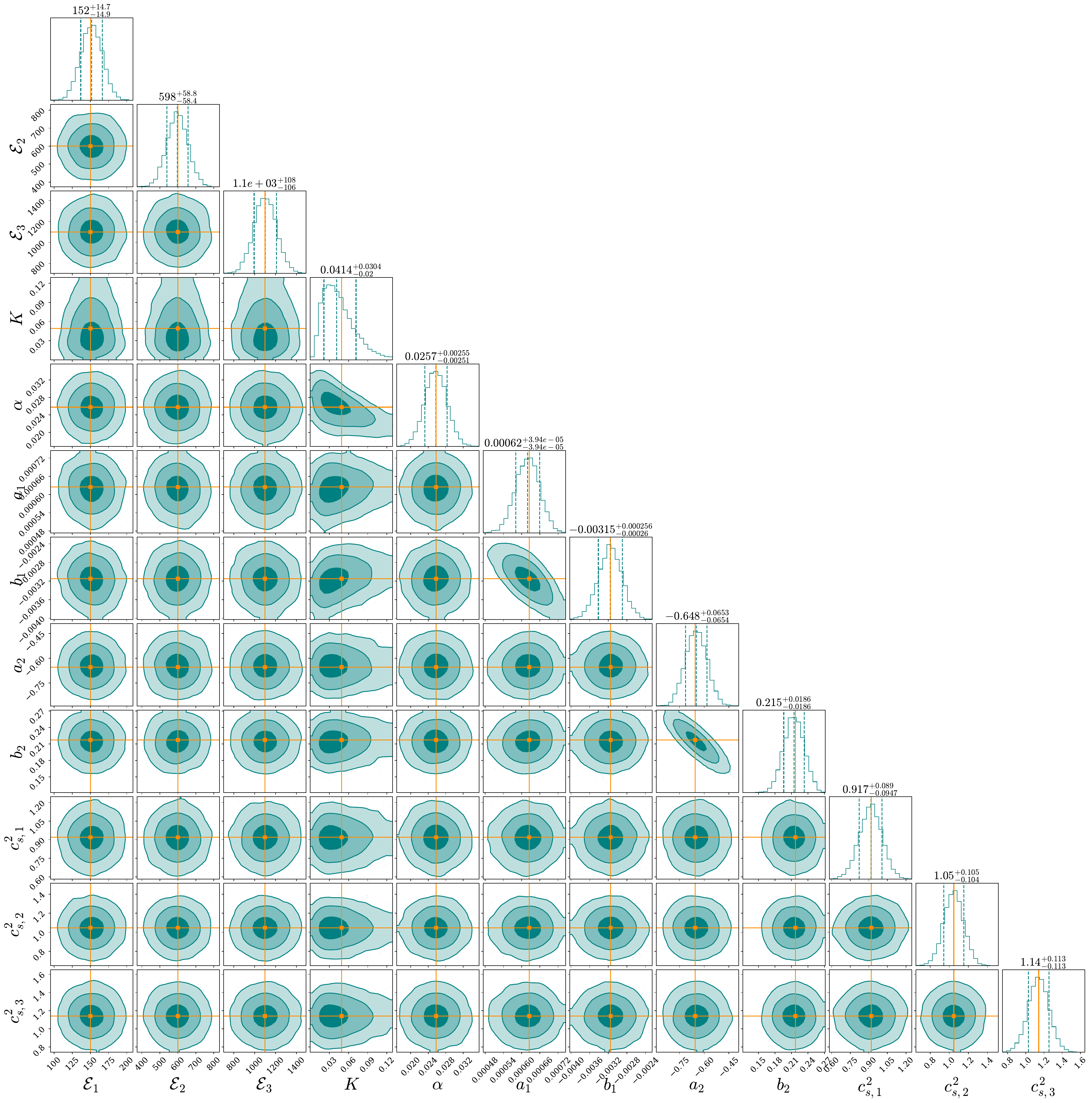}
    \caption{Posterior distributions of the fitting parameters for BSK21 EoS obtained from MM data driven Bayesian inference.}
    \label{fig:BSK21_Posteriwor_MM}
\end{figure}

 Moreover, as discussed before, due to continuity conditions on pressure at various matching points, the parameters $K$ and $b_2$ are not independent parameters and can be derived from the other independent parameters. Hence, it follows that $K$ and $b_2$ are not part of the prior set. At each sample, they are fixed by the pressure-continuity conditions at the bridge intersections, given the sampled values of the other parameters. Hence, the posterior distributions of $K$ and $b_2$ in both Fig. \ref{fig:BSK21_Fit_Posterior} and Fig. \ref{fig:BSK21_Posteriwor_MM} are derived posteriors from the posterior distributions of the independent parameters.
\par The noise driven synthetic EoS generation brings out correlations between various fitting parameters. Several parameter pairs display strong correlations in the BSK21 case, and these correlations have a clear physical origin rather than being generic
statistical artifacts. Within each bridge segment, ${K-\alpha}$ and
$\{a_i, b_i\}$ are anti-correlated, as expected, since both control the same
pressures at the edges of the bridge enforced by the continuity condition:
$a_i$ and $b_i$ can trade off against each other while preserving continuity
at the bridge boundaries. The core sound-speed segments show similar correlations, with $c^2_{s,2}$ and $c^2_{s,3}$ anti-correlated, and $c^2_{s,1}-\epsilon_3$,
$c^2_{s,2}-\epsilon_3$, and $c^2_{s,1}-c^2_{s,2}$ positively correlated.
Moreover, for the EoSs where in the high energy density regimes an
additional parameter $c^2_{s,4}$ is required, $c^2_{s,3}$ and $c^2_{s,4}$
turn out to be negatively correlated. These correlations turn out to be
fairly general for all families of EoSs considered. The other pairs are
usually uncorrelated.
\par Fig.~\ref{fig:BSK21_Posteriwor_MM} shows the posterior of fitting parameters obtained from the MM data driven Bayesian inference, using the maximized log-likelihood in which the MM data, e.g. masses and tidal-deformability measurements (wherever available) of Table.~\ref{tab:Data} are used via Eq. \eqref{eq:logL_total}. For each, only those compact objects are taken into account for which $M \geq 1.25 M_\odot$ and $M < M_{\text{max},\text{eos}}$. As before, $K-\alpha$ and each
$\{a_i, b_i\}$ pair within a bridge segment remains strongly anti-correlated,
consistent with both sets of parameters controlling the same pressure at the
segment boundary under the imposed continuity condition. This is confirmed
explicitly for the second segment as well, where $a_2$ and $b_2$ display the
same tilted, anti-correlated behaviour as $a_1$ and $b_1$.
\par In contrast, the correlations among the core sound-speed parameters
$c^2_{s,1}$, $c^2_{s,2}$, and $c^2_{s,3}$ are no longer visible once MM data are used. This is not because the underlying continuity condition linking the sound-speed segments has changed. Rather, it reflects the substantially
weaker constraints obtained from very limited MM observations, with marginalized uncertainties on $c^2_{s,i}$ roughly an order of magnitude larger than in the synthetic fit. For any particular EoS, the very limited availability of MM data makes the log-likelihood poorly informed. For instance, for $c^2_{s,1}$, the uncertainty around the median of its posterior distribution increases from $\sim 0.01$ from Syn EoS generation to $\sim 0.1$ in MM driven inference. This is also apparent from Fig.~\ref{fig:EoS_APR} and Fig.~\ref{fig:EoS_BSK21}, the band containing the posterior samples obtained from the synthetic data driven analysis is much narrower compared to the band containing the posterior EoS samples obtained from the MM driven likelihood analysis, despite the injection parameters being the same APR reference value. It is expected that, given the data in the synthetic data driven analysis were generated from the true APR EoS itself with prescribed noise, mainly to recover and refine the fitting parameters, the MM framework provides no such assurance.

\par Therefore, any correlation between the sound-speed segments and $\mathcal{E}_3$ is obscured by the width of the posterior distribution rather than being genuinely absent. With such a limited number of MM measurements, the likelihood alone is not informative enough to resolve the correlation structure. We, therefore, do not interpret the disappearance of certain correlations as evidence that the underlying correlations have been lifted by the MM data. In order for the MM data driven likelihood analysis to be perform better in bringing out such correlations between parameters, a more versatile set of MM data is required.

\section{Conclusion}
\label{sec:conclusion}
In this work, we have proposed an analytic functional form capable of describing the EoSs across a wide range of EoS obtained from diverse theoretical considerations, ranging from very soft (e.g. the AP family) to very stiff (e.g. the BSK and SK families) EoSs. The parametrization combines an exponential crust segment, followed by several intermediate bridge segments matched via continuity of pressure at their edges, and the core part described by segment-wise constant speeds of sound.
\par We carry out the fitting in three successive stages: an initial deterministic $\chi^2$-fit to the tabulated EoS, a Bayesian framework using synthetic EoSs generated by injecting Gaussian noise into the fitted EoS to simulate observational uncertainty, and a further Bayesian analysis using an independent log-likelihood using MM data built from measured masses, and tidal deformabilities of compact objects. Across most of the EoS families considered,, the fitting form reproduces both the tabulated EoS profile and the resulting macroscopic observables, e.g. the maximum stellar, stellar radius at the maximum mass, stellar radius for a canonical $M = 1.4~M_\odot$ star and $\Lambda_{1.4}$, to within a few percent of the values obtained directly from the original tabulated EoS. The largest deviations occur in $\Lambda_{1.4}$ because of its strong sensitivity to stellar compactness ($\Lambda \sim C^{-5}$). This level of agreement holds consistently for both soft and stiff EoSs, indicating that the proposed parametrization is not biased toward any particular class of microphysical model and can serve as a general-purpose analytic representation for EoS tables drawn from repositories such as CompOSE and LALSimulation.
\par The posterior distributions of the fitting parameters reveal that they are not mutually independent. Several pairs display strong correlations that arise directly from the continuity conditions imposed at the segment boundaries: whenever two parameters jointly control the pressure at a shared boundary between adjacent segments, they can trade off against one another while leaving the fit to the actual data essentially unchanged, producing a clear anti-correlation between them. This pattern recurs at every segment boundary in the parametrization. These correlations are found to be a fairly generic feature of the parametrization, appearing consistently across the different families of EoSs studied.
\par Comparing the synthetic EoS driven and MM driven posteriors shows that not all of these correlations behave alike under these two conditions. The correlations associated with the low and intermediate density segments persist essentially unchanged in both the Syn and MM analysis. By contrast, the correlations among the high-density (core) segment parameters which are clearly present in the Syntheic data driven fit are not clearly apparent in the MM data driven analysis. This disappearance should not be interpreted as evidence that the underlying correlation has been broken by the MM data; rather, it is a consequence of the substantially wider uncertainties in the posterior distributions of the parameters obtained from very limited number of MM measurements. This is also evident from the observation that the width of the posterior EoS samples obtained from mock EoS generation is much smaller that the posterior EoS samples obtained from MM data driven analysis. With a likelihood this weakly informative, an underlying correlation can easily be masked by the width of the posterior distribution rather than genuinely removed. Distinguishing a breaking of a correlation from an artifact of insufficient data will require Bayesian inference with a larger and more diverse set of MM observations.

\section*{Acknowledgment}

Part of the numerical computations presented in this work was performed on the Pegasus computing cluster at IUCAA, Pune, India, under project No. hpc2502005. AD acknowledges IISERTVM for financial support and for the computational resources required for this research.

\onecolumn

\setlength{\LTcapwidth}{\linewidth}
\begin{longtable}
{@{\extracolsep{\fill}}llrrrrl}
\caption{Comparison of the maximum stellar mass $M_{\text{max}}$, stellar radius at $M_{\text{max}}$, $R_{\text{max}}$, dimensionless tidal deformability at $M = 1.4~M_\odot$, $\Lambda_{1.4}$ and stellar radius at $M = 1.4~M_\odot$, $R_{1.4}$ obtained from the Syn analysis, median EoS of the MM analysis, initially fitted EoS (Fit) obtained from the original EoS (LAL).}\label{tab:macro-comparison}\\
\toprule
EoS & Method & $M_{\max}/M_\odot$ & $R_{\max}$ (km) & $\Lambda_{1.4}$ & $R_{1.4}$ (km) \\
\midrule
\endfirsthead
\toprule
EoS & Method & $M_{\max}/M_\odot$ & $R_{\max}$ (km) & $\Lambda_{1.4}$ & $R_{1.4}$ (km) \\
\midrule
\endhead
\endfoot
\bottomrule
\endlastfoot
\multirow{4}{*}{ALF1} & Syn & 1.496 & 9.26 & 103.07 & 9.98 \\*
 & MM & 1.489 & 9.23 & 99.85 & 9.92 \\*
 & Fit & 1.450 & 9.20 & 99.65 & 9.89 \\*
 & LAL & 1.495 & 9.16 & 91.14 & 9.85 \\
\midrule
\multirow{4}{*}{ALF3} & Syn & 1.478 & 9.57 & 127.81 & 10.42 \\*
 & MM & 1.487 & 9.62 & 133.47 & 10.48 \\*
 & Fit & 1.474 & 9.42 & 115.84 & 10.21 \\*
 & LAL & 1.472 & 9.28 & 122.20 & 10.27 \\
\midrule
\multirow{4}{*}{ALF4} & Syn & 1.947 & 10.94 & 362.99 & 11.68 \\*
 & MM & 1.836 & 10.65 & 311.69 & 11.45 \\*
 & Fit & 1.936 & 10.86 & 347.86 & 11.60 \\*
 & LAL & 1.942 & 10.83 & 329.10 & 11.63 \\
\midrule
\multirow{4}{*}{AP1} & Syn & 1.684 & 8.26 & 69.30 & 9.31 \\*
 & MM & 1.675 & 8.24 & 67.88 & 9.26 \\*
 & Fit & 1.687 & 8.28 & 70.03 & 9.32 \\*
 & LAL & 1.683 & 8.25 & 71.43 & 9.32 \\
\midrule
\multirow{4}{*}{AP2} & Syn & 1.815 & 8.75 & 134.77 & 10.24 \\*
 & MM & 1.809 & 8.74 & 131.81 & 10.20 \\*
 & Fit & 1.817 & 8.73 & 132.97 & 10.11 \\*
 & LAL & 1.807 & 8.69 & 122.89 & 10.14 \\
\midrule
\multirow{4}{*}{AP3} & Syn & 2.405 & 10.81 & 421.59 & 12.17 \\*
 & MM & 2.399 & 10.78 & 422.05 & 12.18 \\*
 & Fit & 2.393 & 10.80 & 429.12 & 12.22 \\*
 & LAL & 2.389 & 10.75 & 417.93 & 12.06 \\
\midrule
\multirow{4}{*}{AP4} & Syn & 2.215 & 10.08 & 310.68 & 11.68 \\*
 & MM & 2.226 & 10.10 & 304.89 & 11.63 \\*
 & Fit & 2.215 & 10.06 & 296.57 & 11.60 \\*
 & LAL & 2.212 & 9.99 & 283.27 & 11.40 \\
\midrule
\multirow{4}{*}{APR} & Syn & 2.189 & 9.96 & 273.74 & 11.34 \\*
 & MM & 2.192 & 9.98 & 275.70 & 11.35 \\*
 & Fit & 2.192 & 9.85 & 280.72 & 11.33 \\*
 & LAL & 2.189 & 9.83 & 274.83 & 11.31 \\
\midrule
\multirow{4}{*}{APR4E} & Syn & 2.169 & 10.22 & 277.15 & 11.32 \\*
 & MM & 2.187 & 10.25 & 278.92 & 11.33 \\*
 & Fit & 2.164 & 10.17 & 257.17 & 11.28 \\*
 & LAL & 2.158 & 10.12 & 263.32 & 11.28 \\
\midrule
\multirow{4}{*}{BSK19} & Syn & 1.859 & 9.14 & 183.03 & 10.75 \\*
 & MM & 1.852 & 9.09 & 172.58 & 10.64 \\*
 & Fit & 1.861 & 9.07 & 183.35 & 10.77 \\*
 & LAL & 1.861 & 9.03 & 183.17 & 10.71 \\
\midrule
\multirow{4}{*}{BSK21} & Syn & 2.290 & 11.13 & 630.81 & 12.69 \\*
 & MM & 2.258 & 10.91 & 557.33 & 12.47 \\*
 & Fit & 2.278 & 11.07 & 604.87 & 12.63 \\*
 & LAL & 2.275 & 11.05 & 576.36 & 12.56 \\
\midrule
\multirow{4}{*}{FPS} & Syn & 1.798 & 9.40 & 191.27 & 11.05 \\*
 & MM & 1.786 & 9.37 & 183.50 & 10.96 \\*
 & Fit & 1.800 & 9.40 & 191.28 & 10.92 \\*
 & LAL & 1.801 & 9.33 & 196.77 & 10.83 \\
\midrule
\multirow{4}{*}{H4} & Syn & 2.046 & 11.82 & 974.30 & 13.74 \\*
 & MM & 2.031 & 11.70 & 906.93 & 13.53 \\*
 & Fit & 2.036 & 11.73 & 927.48 & 13.63 \\*
 & LAL & 2.031 & 11.69 & 932.41 & 13.64 \\
\midrule
\multirow{4}{*}{KDE0V} & Syn & 1.960 & 9.69 & 275.66 & 11.59 \\*
 & MM & 1.953 & 9.65 & 264.95 & 11.51 \\*
 & Fit & 1.958 & 9.65 & 272.34 & 11.55 \\*
 & LAL & 1.960 & 9.66 & 253.71 & 11.40 \\
\midrule
\multirow{4}{*}{KDE0V1} & Syn & 1.971 & 9.79 & 312.69 & 11.66 \\*
 & MM & 1.957 & 9.71 & 291.68 & 11.54 \\*
 & Fit & 1.967 & 9.76 & 294.07 & 11.65 \\*
 & LAL & 1.969 & 9.79 & 281.30 & 11.60 \\
\midrule
\multirow{4}{*}{SK255} & Syn & 2.140 & 10.84 & 599.31 & 13.48 \\*
 & MM & 2.150 & 10.72 & 619.32 & 12.71 \\*
 & Fit & 2.138 & 10.64 & 566.62 & 12.72 \\*
 & LAL & 2.143 & 10.85 & 613.72 & 13.02 \\
\midrule
\multirow{4}{*}{SK272} & Syn & 2.236 & 11.11 & 735.14 & 13.24 \\*
 & MM & 2.205 & 10.96 & 667.16 & 13.18 \\*
 & Fit & 2.228 & 11.01 & 696.25 & 13.33 \\*
 & LAL & 2.231 & 11.03 & 668.59 & 13.29 \\
\midrule
\multirow{4}{*}{SKA} & Syn & 2.214 & 10.93 & 688.55 & 13.02 \\*
 & MM & 2.169 & 10.67 & 591.99 & 12.84 \\*
 & Fit & 2.208 & 10.93 & 663.01 & 13.12 \\*
 & LAL & 2.208 & 10.92 & 590.33 & 12.89 \\
\midrule
\multirow{4}{*}{SKB} & Syn & 2.193 & 10.61 & 548.73 & 12.03 \\*
 & MM & 2.194 & 10.59 & 538.20 & 12.05 \\*
 & Fit & 2.184 & 10.50 & 489.88 & 12.06 \\*
 & LAL & 2.187 & 10.67 & 450.66 & 12.18 \\
\midrule
\multirow{4}{*}{SKI2} & Syn & 2.173 & 11.19 & 898.33 & 13.29 \\*
 & MM & 2.137 & 11.63 & 774.32 & 12.95 \\*
 & Fit & 2.164 & 11.60 & 843.90 & 13.33 \\*
 & LAL & 2.162 & 11.10 & 801.54 & 13.46 \\
\midrule
\multirow{4}{*}{SKI3} & Syn & 2.247 & 11.44 & 935.38 & 13.93 \\*
 & MM & 2.196 & 11.05 & 777.00 & 13.41 \\*
 & Fit & 2.237 & 11.30 & 865.54 & 13.74 \\*
 & LAL & 2.239 & 11.29 & 815.22 & 13.53 \\
\midrule
\multirow{4}{*}{SKI4} & Syn & 2.178 & 10.76 & 551.45 & 12.57 \\*
 & MM & 2.181 & 10.72 & 536.83 & 12.51 \\*
 & Fit & 2.170 & 10.63 & 514.90 & 12.42 \\*
 & LAL & 2.169 & 10.64 & 478.11 & 12.35 \\
\midrule
\multirow{4}{*}{SKI5} & Syn & 2.255 & 11.39 & 1103.68 & 13.60 \\*
 & MM & 2.170 & 10.89 & 788.36 & 12.90 \\*
 & Fit & 2.241 & 11.35 & 1011.25 & 13.44 \\*
 & LAL & 2.239 & 11.39 & 1044.11 & 14.06 \\
\midrule
\multirow{4}{*}{SKI6} & Syn & 2.193 & 10.80 & 551.75 & 12.53 \\*
 & MM & 2.185 & 10.75 & 537.10 & 12.47 \\*
 & Fit & 2.189 & 10.74 & 537.10 & 12.52 \\*
 & LAL & 2.189 & 10.74 & 521.02 & 12.47 \\
\midrule
\multirow{4}{*}{SKMP} & Syn & 2.116 & 10.58 & 538.11 & 12.50 \\*
 & MM & 2.120 & 10.56 & 533.38 & 12.46 \\*
 & Fit & 2.110 & 10.50 & 511.59 & 12.39 \\*
 & LAL & 2.107 & 10.53 & 500.18 & 12.47 \\
\midrule
\multirow{4}{*}{SKOP} & Syn & 1.979 & 10.10 & 398.93 & 11.99 \\*
 & MM & 2.000 & 10.17 & 415.76 & 12.07 \\*
 & Fit & 1.976 & 10.07 & 384.60 & 12.06 \\*
 & LAL & 1.972 & 10.10 & 380.98 & 12.11 \\
\midrule
\multirow{4}{*}{SLY} & Syn & 2.058 & 10.08 & 342.78 & 11.92 \\*
 & MM & 2.034 & 9.91 & 300.29 & 11.70 \\*
 & Fit & 2.054 & 10.05 & 340.79 & 11.90 \\*
 & LAL & 2.050 & 9.98 & 338.51 & 11.73 \\
\midrule
\multirow{4}{*}{SLY2} & Syn & 2.052 & 10.14 & 353.08 & 11.90 \\*
 & MM & 2.021 & 9.93 & 305.26 & 11.78 \\*
 & Fit & 2.049 & 9.95 & 327.04 & 11.71 \\*
 & LAL & 2.053 & 10.01 & 321.88 & 11.76 \\
\midrule
\multirow{4}{*}{SLY4} & Syn & 2.055 & 10.10 & 333.19 & 12.01 \\*
 & MM & 2.060 & 10.09 & 329.86 & 11.98 \\*
 & Fit & 2.052 & 10.04 & 332.65 & 11.87 \\*
 & LAL & 2.050 & 9.98 & 338.53 & 11.73 \\
\midrule
\multirow{4}{*}{WFF1} & Syn & 2.139 & 9.49 & 176.84 & 10.51 \\*
 & MM & 2.154 & 9.52 & 180.11 & 10.54 \\*
 & Fit & 2.141 & 9.54 & 183.44 & 10.47 \\*
 & LAL & 2.134 & 9.48 & 177.97 & 10.38 \\
\midrule
\multirow{4}{*}{WFF2} & Syn & 2.200 & 9.82 & 246.66 & 11.19 \\*
 & MM & 2.218 & 9.91 & 258.71 & 11.28 \\*
 & Fit & 2.204 & 9.74 & 255.17 & 11.18 \\*
 & LAL & 2.199 & 9.70 & 248.04 & 11.13 \\
\midrule
\multirow{4}{*}{WFF3} & Syn & 1.844 & 9.58 & 221.17 & 11.03 \\*
 & MM & 1.823 & 9.48 & 198.88 & 10.91 \\*
 & Fit & 1.842 & 9.40 & 212.82 & 10.95 \\*
 & LAL & 1.844 & 9.37 & 186.85 & 10.89 \\
\end{longtable}

\onecolumn
\footnotesize
	
\renewcommand{\arraystretch}{1.1}
\setlength{\tabcolsep}{1pt}

\begin{longtable}[c]{
@{}
>{\centering\arraybackslash}p{0.06\textwidth}
>{\centering\arraybackslash}p{0.06\textwidth}
>{\centering\arraybackslash}p{0.16\textwidth}
>{\centering\arraybackslash}p{0.14\textwidth}
>{\centering\arraybackslash}p{0.25\textwidth}
>{\centering\arraybackslash}p{0.18\textwidth}
|
>{\centering\arraybackslash}p{0.13\textwidth}
@{}
}

\caption{Parameters of the piecewise-continuous representation for the considered EoS. For each EoS, the \textbf{Fit} row lists the parameters obtained from the initial deterministic fit to the tabulated EoS. The \textbf{Syn} and \textbf{MM} rows report the posterior median values obtained, respectively, from synthetic noisy EoS data and from using NS MM observations. Here, $\epsilon_i=(\epsilon_{1},\epsilon_{2},\epsilon_{3})$ denotes the segment-matching energy densities, $(a_i,b_i)$ are the bridge parameters and $c_s^2$ gives the squared sound speed in each core segment. The bridge exponents $(\alpha_i,\beta_i)$ are fixed for each EoS, for both approaches.}
\label{dataset_table}\\

\toprule
\toprule
\textbf{EoS}
&
\textbf{Method}
&
$\mathcal{E}_i~(\mathrm{MeV/fm^3})$
&
$(K,\alpha)\times10^{-2}$
&
$(a_i,b_i)$
&
$c_s^2$
&
\begin{tabular}{@{}c@{}}
$(\alpha_i,\beta_i)$\\
\textit{(fixed)}
\end{tabular}
\\
\midrule
\endfirsthead

\toprule
\textbf{EoS}
&
\textbf{Method}
&
$\mathcal{E}_i~(\mathrm{MeV/fm^3})$
&
$(K,\alpha)\times10^{-2}$
&
$(a_i,b_i)$
&
$c_s^2$
&
\begin{tabular}{@{}c@{}}
$(\alpha_i,\beta_i)$\\
\textit{(fixed)}
\end{tabular}
\\
\midrule
\endhead

\endfoot

\bottomrule
\endlastfoot

\midrule

\multirow{3}{*}{\textbf{ALF1}}
&
\textbf{Fit}
&
\begin{tabular}{@{}c@{}}
$(70.000,\;220.000,$\\
$450.000)$
\end{tabular}
&
$(11.167,\;1.935)$
&
\begin{tabular}{@{}c@{}}
$(1.407\times10^{-7},\;1.271\times10^{-3})$\\
$(2.107\times10^{-19},\;0.053)$
\end{tabular}
&
$(0.296,\;0.318,\;0.320)$
&
\begin{tabular}{@{}c@{}}
$(3.200,\;1.200)$\\
$(7.500,\;0.850)$
\end{tabular}
\\[3mm]

&
\textbf{Syn}
&
\begin{tabular}{@{}c@{}}
$(69.930,\;224.253,$\\
$444.048)$
\end{tabular}
&
$(11.135,\;1.942)$
&
\begin{tabular}{@{}c@{}}
$(1.418\times10^{-7},\;1.273\times10^{-3})$\\
$(2.151\times10^{-19},\;0.055)$
\end{tabular}
&
$(0.291,\;0.320,\;0.321)$
&
\\[3mm]

&
\textbf{MM}
&
\begin{tabular}{@{}c@{}}
$(69.982,\;221.231,$\\
$448.788)$
\end{tabular}
&
$(11.128,\;1.938)$
&
\begin{tabular}{@{}c@{}}
$(1.408\times10^{-7},\;1.269\times10^{-3})$\\
$(2.107\times10^{-19},\;0.053)$
\end{tabular}
&
$(0.290,\;0.316,\;0.321)$
&
\\[3mm]

\midrule

\multirow{3}{*}{\textbf{ALF3}}
&
\textbf{Fit}
&
\begin{tabular}{@{}c@{}}
$(100.000,\;355.000,$\\
$600.000)$
\end{tabular}
&
$(10.160,\;2.061)$
&
\begin{tabular}{@{}c@{}}
$(5.827\times10^{-8},\;1.611\times10^{-3})$\\
$(8.155\times10^{-11},\;0.091)$
\end{tabular}
&
$(0.278,\;0.318,\;0.320)$
&
\begin{tabular}{@{}c@{}}
$(3.350,\;1.200)$\\
$(4.200,\;0.900)$
\end{tabular}
\\[3mm]

&
\textbf{Syn}
&
\begin{tabular}{@{}c@{}}
$(99.908,\;367.143,$\\
$596.360)$
\end{tabular}
&
$(10.139,\;2.064)$
&
\begin{tabular}{@{}c@{}}
$(5.851\times10^{-8},\;1.609\times10^{-3})$\\
$(8.020\times10^{-11},\;0.099)$
\end{tabular}
&
$(0.272,\;0.321,\;0.321)$
&
\\[3mm]

&
\textbf{MM}
&
\begin{tabular}{@{}c@{}}
$(100.127,\;363.609,$\\
$598.631)$
\end{tabular}
&
$(10.234,\;2.063)$
&
\begin{tabular}{@{}c@{}}
$(5.944\times10^{-8},\;1.613\times10^{-3})$\\
$(8.196\times10^{-11},\;0.098)$
\end{tabular}
&
$(0.275,\;0.317,\;0.320)$
&
\\[3mm]

\midrule

\multirow{3}{*}{\textbf{ALF4}}
&
\textbf{Fit}
&
\begin{tabular}{@{}c@{}}
$(90.000,\;290.000,$\\
$860.000)$
\end{tabular}
&
$(8.796,\;2.218)$
&
\begin{tabular}{@{}c@{}}
$(7.747\times10^{-8},\;1.544\times10^{-3})$\\
$(0.049,\;-2.026)$
\end{tabular}
&
$(0.300,\;0.252,\;0.318)$
&
\begin{tabular}{@{}c@{}}
$(3.300,\;1.200)$\\
$(1.330,\;0.650)$
\end{tabular}
\\[3mm]

&
\textbf{Syn}
&
\begin{tabular}{@{}c@{}}
$(90.093,\;284.413,$\\
$848.419)$
\end{tabular}
&
$(8.826,\;2.219)$
&
\begin{tabular}{@{}c@{}}
$(7.825\times10^{-8},\;1.546\times10^{-3})$\\
$(0.049,\;-2.013)$
\end{tabular}
&
$(0.297,\;0.256,\;0.317)$
&
\\[3mm]

&
\textbf{MM}
&
\begin{tabular}{@{}c@{}}
$(89.899,\;297.731,$\\
$845.166)$
\end{tabular}
&
$(8.772,\;2.217)$
&
\begin{tabular}{@{}c@{}}
$(7.667\times10^{-8},\;1.544\times10^{-3})$\\
$(0.045,\;-1.833)$
\end{tabular}
&
$(0.297,\;0.252,\;0.318)$
&
\\[3mm]


\midrule

\multirow{3}{*}{\textbf{AP1}}
&
\textbf{Fit}
&
\begin{tabular}{@{}c@{}}
$(80.000,\;200.000,$\\
$1062.000)$
\end{tabular}
&
$(12.166,\;1.940)$
&
\begin{tabular}{@{}c@{}}
$(3.570\times10^{-12},\;2.872\times10^{-3})$\\
$(3.146\times10^{-5},\;-1.472\times10^{-4})$
\end{tabular}
&
$(0.651,\;0.790,\;0.736)$
&
\begin{tabular}{@{}c@{}}
$(4.950,\;1.150)$\\
$(2.350,\;2.000)$
\end{tabular}
\\[3mm]

&
\textbf{Syn}
&
\begin{tabular}{@{}c@{}}
$(80.071,\;197.043,$\\
$1126.180)$
\end{tabular}
&
$(12.375,\;1.942)$
&
\begin{tabular}{@{}c@{}}
$(3.621\times10^{-12},\;2.928\times10^{-3})$\\
$(3.086\times10^{-5},\;-1.420\times10^{-4})$
\end{tabular}
&
$(0.665,\;0.801,\;0.728)$
&
\\[3mm]

&
\textbf{MM}
&
\begin{tabular}{@{}c@{}}
$(79.941,\;200.667,$\\
$1072.872)$
\end{tabular}
&
$(12.156,\;1.940)$
&
\begin{tabular}{@{}c@{}}
$(3.567\times10^{-12},\;2.869\times10^{-3})$\\
$(3.092\times10^{-5},\;-1.440\times10^{-4})$
\end{tabular}
&
$(0.637,\;0.785,\;0.736)$
&
\\[3mm]

\midrule

\multirow{3}{*}{\textbf{AP2}}
&
\textbf{Fit}
&
\begin{tabular}{@{}c@{}}
$(35.000,\;250.000,$\\
$1100.000)$
\end{tabular}
&
$(4.244,\;3.821)$
&
\begin{tabular}{@{}c@{}}
$(1.324\times10^{-7},\;1.537\times10^{-3})$\\
$(-1.890\times10^{-3},\;7.208\times10^{-4})$
\end{tabular}
&
$(0.766,\;0.967,\;1.117)$
&
\begin{tabular}{@{}c@{}}
$(3.150,\;1.200)$\\
$(1.800,\;2.000)$
\end{tabular}
\\[3mm]

&
\textbf{Syn}
&
\begin{tabular}{@{}c@{}}
$(35.069,\;247.567,$\\
$1141.350)$
\end{tabular}
&
$(4.280,\;3.820)$
&
\begin{tabular}{@{}c@{}}
$(1.358\times10^{-7},\;1.548\times10^{-3})$\\
$(-1.853\times10^{-3},\;7.110\times10^{-4})$
\end{tabular}
&
$(0.779,\;0.985,\;1.113)$
&
\\[3mm]

&
\textbf{MM}
&
\begin{tabular}{@{}c@{}}
$(35.058,\;249.707,$\\
$1117.108)$
\end{tabular}
&
$(4.266,\;3.815)$
&
\begin{tabular}{@{}c@{}}
$(1.326\times10^{-7},\;1.542\times10^{-3})$\\
$(-1.869\times10^{-3},\;7.140\times10^{-4})$
\end{tabular}
&
$(0.754,\;0.962,\;1.116)$
&
\\[3mm]

\midrule

\multirow{3}{*}{\textbf{AP3}}
&
\textbf{Fit}
&
\begin{tabular}{@{}c@{}}
$(200.000,\;320.000,$\\
$1000.000)$
\end{tabular}
&
$(16.126,\;1.717)$
&
\begin{tabular}{@{}c@{}}
$(1.039\times10^{-8},\;2.506\times10^{-3})$\\
$(2.902\times10^{-2},\;-4.953\times10^{-2})$
\end{tabular}
&
$(1.249,\;1.403,\;1.481)$
&
\begin{tabular}{@{}c@{}}
$(3.700,\;1.200)$\\
$(2.000,\;1.700)$
\end{tabular}
\\[3mm]

&
\textbf{Syn}
&
\begin{tabular}{@{}c@{}}
$(200.031,\;321.031,$\\
$735.936)$
\end{tabular}
&
$(16.194,\;1.718)$
&
\begin{tabular}{@{}c@{}}
$(1.050\times10^{-8},\;2.505\times10^{-3})$\\
$(1.390\times10^{-3},\;-6.630\times10^{-3})$
\end{tabular}
&
$(1.120,\;1.390,\;1.474)$
&
\\[3mm]

&
\textbf{MM}
&
\begin{tabular}{@{}c@{}}
$(199.975,\;319.731,$\\
$803.324)$
\end{tabular}
&
$(16.435,\;1.712)$
&
\begin{tabular}{@{}c@{}}
$(1.050\times10^{-8},\;2.510\times10^{-3})$\\
$(1.388\times10^{-3},\;-6.618\times10^{-3})$
\end{tabular}
&
$(1.156,\;1.385,\;1.476)$
&
\\[3mm]

\midrule

\multirow{3}{*}{\textbf{AP4}}
&
\textbf{Fit}
&
\begin{tabular}{@{}c@{}}
$(180.000,\;350.000,$\\
$1200.000)$
\end{tabular}
&
$(17.631,\;1.625)$
&
\begin{tabular}{@{}c@{}}
$(1.295\times10^{-8},\;2.762\times10^{-3})$\\
$(4.014\times10^{-4},\;-0.070)$
\end{tabular}
&
$(1.275,\;1.430,\;1.522)$
&
\begin{tabular}{@{}c@{}}
$(3.600,\;1.200)$\\
$(2.050,\;1.100)$
\end{tabular}
\\[3mm]

&
\textbf{Syn}
&
\begin{tabular}{@{}c@{}}
$(179.801,\;338.916,$\\
$1249.748)$
\end{tabular}
&
$(18.149,\;1.626)$
&
\begin{tabular}{@{}c@{}}
$(1.357\times10^{-8},\;2.787\times10^{-3})$\\
$(3.942\times10^{-4},\;-0.066)$
\end{tabular}
&
$(1.286,\;1.455,\;1.509)$
&
\\[3mm]

&
\textbf{MM}
&
\begin{tabular}{@{}c@{}}
$(180.051,\;344.851,$\\
$1200.824)$
\end{tabular}
&
$(17.998,\;1.621)$
&
\begin{tabular}{@{}c@{}}
$(1.322\times10^{-8},\;2.775\times10^{-3})$\\
$(4.060\times10^{-4},\;-0.070)$
\end{tabular}
&
$(1.272,\;1.430,\;1.520)$
&
\\[3mm]

\midrule

\multirow{3}{*}{\textbf{APR}}
&
\textbf{Fit}
&
\begin{tabular}{@{}c@{}}
$(70.000,\;320.000,$\\
$900.000)$
\end{tabular}
&
$(10.513,\;2.018)$
&
\begin{tabular}{@{}c@{}}
$(1.407\times10^{-7},\;1.306\times10^{-3})$\\
$(3.398\times10^{-5},\;-3.323\times10^{-3})$
\end{tabular}
&
$(1.046,\;1.274,\;1.439)$
&
\begin{tabular}{@{}c@{}}
$(3.200,\;1.200)$\\
$(2.400,\;1.500)$
\end{tabular}
\\[3mm]

&
\textbf{Syn}
&
\begin{tabular}{@{}c@{}}
$(70.035,\;318.163,$\\
$849.833)$
\end{tabular}
&
$(10.584,\;2.020)$
&
\begin{tabular}{@{}c@{}}
$(1.434\times10^{-7},\;1.309\times10^{-3})$\\
$(3.375\times10^{-5},\;-3.225\times10^{-3})$
\end{tabular}
&
$(1.020,\;1.275,\;1.435)$
&
\\[3mm]

&
\textbf{MM}
&
\begin{tabular}{@{}c@{}}
$(70.012,\;317.186,$\\
$903.695)$
\end{tabular}
&
$(10.647,\;2.014)$
&
\begin{tabular}{@{}c@{}}
$(1.435\times10^{-7},\;1.310\times10^{-3})$\\
$(3.410\times10^{-5},\;-3.283\times10^{-3})$
\end{tabular}
&
$(1.036,\;1.272,\;1.435)$
&
\\[3mm]

\midrule

\multirow{3}{*}{\textbf{APR4$\_$EPP}}
&
\textbf{Fit}
&
\begin{tabular}{@{}c@{}}
$(70.000,\;500.000,$\\
$1190.000)$
\end{tabular}
&
$(14.826,\;1.595)$
&
\begin{tabular}{@{}c@{}}
$(5.649\times10^{-8},\;1.337\times10^{-3})$\\
$(-0.451,\;0.106)$
\end{tabular}
&
$(0.688,\;0.721,\;0.748)$
&
\begin{tabular}{@{}c@{}}
$(3.250,\;1.200)$\\
$(1.150,\;1.400)$
\end{tabular}
\\[3mm]

&
\textbf{Syn}
&
\begin{tabular}{@{}c@{}}
$(69.825,\;497.522,$\\
$1190.551)$
\end{tabular}
&
$(15.011,\;1.597)$
&
\begin{tabular}{@{}c@{}}
$(1.101\times10^{-7},\;1.221\times10^{-3})$\\
$(-0.433,\;0.103)$
\end{tabular}
&
$(0.690,\;0.730,\;0.741)$
&
\\[3mm]

&
\textbf{MM}
&
\begin{tabular}{@{}c@{}}
$(70.048,\;493.478,$\\
$1188.877)$
\end{tabular}
&
$(15.028,\;1.600)$
&
\begin{tabular}{@{}c@{}}
$(1.105\times10^{-7},\;1.224\times10^{-3})$\\
$(-0.451,\;0.107)$
\end{tabular}
&
$(0.687,\;0.720,\;0.747)$
&
\\[3mm]

\midrule


\multirow{3}{*}{\textbf{BSK19}}
&
\textbf{Fit}
&
\begin{tabular}{@{}c@{}}
$(175.000,\;600.000,$\\
$1200.000)$
\end{tabular}
&
$(13.538,\;1.738)$
&
\begin{tabular}{@{}c@{}}
$(1.024\times10^{-5},\;-2.233\times10^{-4})$\\
$(6.206\times10^{-3},\;-0.010)$
\end{tabular}
&
$(0.785,\;0.928,\;1.058)$
&
\begin{tabular}{@{}c@{}}
$(2.500,\;1.700)$\\
$(1.800,\;1.700)$
\end{tabular}
\\[3mm]

&
\textbf{Syn}
&
\begin{tabular}{@{}c@{}}
$(160.036,\;592.865,$\\
$1242.173)$
\end{tabular}
&
$(14.064,\;1.736)$
&
\begin{tabular}{@{}c@{}}
$(1.049\times10^{-5},\;-2.287\times10^{-4})$\\
$(5.862\times10^{-3},\;-0.010)$
\end{tabular}
&
$(0.795,\;0.945,\;1.051)$
&
\\[3mm]

&
\textbf{MM}
&
\begin{tabular}{@{}c@{}}
$(159.880,\;595.957,$\\
$1208.182)$
\end{tabular}
&
$(13.563,\;1.736)$
&
\begin{tabular}{@{}c@{}}
$(1.026\times10^{-5},\;-2.290\times10^{-4})$\\
$(6.111\times10^{-3},\;-0.010)$
\end{tabular}
&
$(0.776,\;0.925,\;1.056)$
&
\\[3mm]

\midrule

\multirow{3}{*}{\textbf{BSK21}}
&
\textbf{Fit}
&
\begin{tabular}{@{}c@{}}
$(150.000,\;600.000,$\\
$1100.000)$
\end{tabular}
&
$(4.922,\;2.578)$
&
\begin{tabular}{@{}c@{}}
$(6.248\times10^{-4},\;-3.148\times10^{-3})$\\
$(-0.655,\;0.217)$
\end{tabular}
&
$(0.920,\;1.048,\;1.141)$
&
\begin{tabular}{@{}c@{}}
$(2.050,\;1.700)$\\
$(1.100,\;1.300)$
\end{tabular}
\\[3mm]

&
\textbf{Syn}
&
\begin{tabular}{@{}c@{}}
$(149.878,\;592.597,$\\
$1170.061)$
\end{tabular}
&
$(5.496,\;2.582)$
&
\begin{tabular}{@{}c@{}}
$(6.300\times10^{-4},\;-3.122\times10^{-3})$\\
$(-0.620,\;0.208)$
\end{tabular}
&
$(0.937,\;1.068,\;1.135)$
&
\\[3mm]

&
\textbf{MM}
&
\begin{tabular}{@{}c@{}}
$(151.621,\;597.791,$\\
$1101.595)$
\end{tabular}
&
$(4.618,\;2.574)$
&
\begin{tabular}{@{}c@{}}
$(6.197\times10^{-4},\;-3.153\times10^{-3})$\\
$(-0.648,\;0.215)$
\end{tabular}
&
$(0.917,\;1.047,\;1.142)$
&
\\[3mm]

\midrule


\multirow{3}{*}{\textbf{FPS}}
&
\textbf{Fit}
&
\begin{tabular}{@{}c@{}}
$(90.000,\;500.000,$\\
$900.000)$
\end{tabular}
&
$(10.546,\;1.691)$
&
\begin{tabular}{@{}c@{}}
$(5.870\times10^{-6},\;-4.069\times10^{-5})$\\
$(1.956\times10^{-3},\;-0.015)$
\end{tabular}
&
$(0.547,\;0.672,\;0.779)$
&
\begin{tabular}{@{}c@{}}
$(2.600,\;2.000)$\\
$(1.800,\;1.400)$
\end{tabular}
\\[3mm]

&
\textbf{Syn}
&
\begin{tabular}{@{}c@{}}
$(90.095,\;494.484,$\\
$845.510)$
\end{tabular}
&
$(10.895,\;1.691)$
&
\begin{tabular}{@{}c@{}}
$(5.965\times10^{-6},\;-4.063\times10^{-5})$\\
$(1.900\times10^{-3},\;-0.014)$
\end{tabular}
&
$(0.532,\;0.671,\;0.776)$
&
\\[3mm]

&
\textbf{MM}
&
\begin{tabular}{@{}c@{}}
$(90.328,\;497.302,$\\
$906.395)$
\end{tabular}
&
$(10.557,\;1.695)$
&
\begin{tabular}{@{}c@{}}
$(5.873\times10^{-6},\;-4.069\times10^{-5})$\\
$(1.924\times10^{-3},\;-0.015)$
\end{tabular}
&
$(0.533,\;0.668,\;0.782)$
&
\\[3mm]

\midrule


\multirow{3}{*}{\textbf{H4}}
&
\textbf{Fit}
&
\begin{tabular}{@{}c@{}}
$(75.000,\;300.000,$\\
$800.000)$
\end{tabular}
&
$(0.847,\;5.552)$
&
\begin{tabular}{@{}c@{}}
$(1.182\times10^{-5},\;-5.393\times10^{-4})$\\
$(0.028,\;-0.083)$
\end{tabular}
&
$(0.444,\;0.430,\;0.443)$
&
\begin{tabular}{@{}c@{}}
$(2.600,\;1.500)$\\
$(1.450,\;1.200)$
\end{tabular}
\\[3mm]

&
\textbf{Syn}
&
\begin{tabular}{@{}c@{}}
$(75.023,\;300.486,$\\
$779.443)$
\end{tabular}
&
$(0.866,\;5.557)$
&
\begin{tabular}{@{}c@{}}
$(1.201\times10^{-5},\;-5.386\times10^{-4})$\\
$(0.027,\;-0.082)$
\end{tabular}
&
$(0.436,\;0.435,\;0.442)$
&
\\[3mm]

&
\textbf{MM}
&
\begin{tabular}{@{}c@{}}
$(75.168,\;303.625,$\\
$798.978)$
\end{tabular}
&
$(0.798,\;5.587)$
&
\begin{tabular}{@{}c@{}}
$(1.161\times10^{-5},\;-5.406\times10^{-4})$\\
$(0.028,\;-0.085)$
\end{tabular}
&
$(0.444,\;0.429,\;0.442)$
&
\\[3mm]

\midrule

\multirow{3}{*}{\textbf{KDE0V}}
&
\textbf{Fit}
&
\begin{tabular}{@{}c@{}}
$(40.000,\;500.000,$\\
$1300.000)$
\end{tabular}
&
$(5.895,\;3.247)$
&
\begin{tabular}{@{}c@{}}
$(3.302\times10^{-6},\;1.103\times10^{-3})$\\
$(1.216\times10^{-3},\;-0.066)$
\end{tabular}
&
$(0.882,\;0.999,\;1.089)$
&
\begin{tabular}{@{}c@{}}
$(2.800,\;1.200)$\\
$(1.850,\;1.100)$
\end{tabular}
\\[3mm]

&
\textbf{Syn}
&
\begin{tabular}{@{}c@{}}
$(39.994,\;493.906,$\\
$1321.552)$
\end{tabular}
&
$(5.972,\;3.245)$
&
\begin{tabular}{@{}c@{}}
$(1.612\times10^{-6},\;1.310\times10^{-3})$\\
$(1.188\times10^{-3},\;-0.061)$
\end{tabular}
&
$(0.888,\;1.015,\;1.078)$
&
\\[3mm]

&
\textbf{MM}
&
\begin{tabular}{@{}c@{}}
$(39.998,\;496.920,$\\
$1300.497)$
\end{tabular}
&
$(5.932,\;3.245)$
&
\begin{tabular}{@{}c@{}}
$(1.578\times10^{-6},\;1.310\times10^{-3})$\\
$(1.191\times10^{-3},\;-0.062)$
\end{tabular}
&
$(0.878,\;0.997,\;1.089)$
&
\\[3mm]

\midrule

\multirow{3}{*}{\textbf{KDE0V1}}
&
\textbf{Fit}
&
\begin{tabular}{@{}c@{}}
$(35.000,\;500.000,$\\
$1300.000)$
\end{tabular}
&
$(4.496,\;3.881)$
&
\begin{tabular}{@{}c@{}}
$(3.413\times10^{-6},\;1.165\times10^{-3})$\\
$(1.797\times10^{-3},\;-0.074)$
\end{tabular}
&
$(0.861,\;0.969,\;1.060)$
&
\begin{tabular}{@{}c@{}}
$(2.680,\;1.200)$\\
$(1.800,\;1.100)$
\end{tabular}
\\[3mm]

&
\textbf{Syn}
&
\begin{tabular}{@{}c@{}}
$(35.015,\;494.978,$\\
$1323.993)$
\end{tabular}
&
$(4.560,\;3.876)$
&
\begin{tabular}{@{}c@{}}
$(3.529\times10^{-6},\;1.164\times10^{-3})$\\
$(1.755\times10^{-3},\;-0.069)$
\end{tabular}
&
$(0.867,\;0.985,\;1.049)$
&
\\[3mm]

&
\textbf{MM}
&
\begin{tabular}{@{}c@{}}
$(34.904,\;499.205,$\\
$1304.644)$
\end{tabular}
&
$(4.516,\;3.871)$
&
\begin{tabular}{@{}c@{}}
$(3.411\times10^{-6},\;1.164\times10^{-3})$\\
$(1.757\times10^{-3},\;-0.071)$
\end{tabular}
&
$(0.855,\;0.964,\;1.060)$
&
\\[3mm]

\midrule


\multirow{3}{*}{\textbf{SK255}}
&
\textbf{Fit}
&
\begin{tabular}{@{}c@{}}
$(30.000,\;500.000,$\\
$1150.000)$
\end{tabular}
&
$(1.098,\;6.976)$
&
\begin{tabular}{@{}c@{}}
$(1.042\times10^{-5},\;4.081\times10^{-4})$\\
$(-0.118,\;0.070)$
\end{tabular}
&
$(0.859,\;0.977,\;1.065)$
&
\begin{tabular}{@{}c@{}}
$(2.500,\;1.100)$\\
$(1.400,\;1.500)$
\end{tabular}
\\[3mm]

&
\textbf{Syn}
&
\begin{tabular}{@{}c@{}}
$(29.962,\;487.206,$\\
$1151.640)$
\end{tabular}
&
$(1.125,\;6.979)$
&
\begin{tabular}{@{}c@{}}
$(1.465\times10^{-5},\;1.855\times10^{-4})$\\
$(-0.112,\;0.068)$
\end{tabular}
&
$(0.862,\;0.991,\;1.057)$
&
\\[3mm]

&
\textbf{MM}
&
\begin{tabular}{@{}c@{}}
$(29.963,\;518.488,$\\
$1139.075)$
\end{tabular}
&
$(1.121,\;6.972)$
&
\begin{tabular}{@{}c@{}}
$(1.455\times10^{-5},\;1.861\times10^{-4})$\\
$(-0.114,\;0.068)$
\end{tabular}
&
$(0.856,\;0.978,\;1.065)$
&
\\[3mm]

\midrule

\multirow{3}{*}{\textbf{SK272}}
&
\textbf{Fit}
&
\begin{tabular}{@{}c@{}}
$(35.000,\;500.000,$\\
$1250.000)$
\end{tabular}
&
$(1.741,\;5.862)$
&
\begin{tabular}{@{}c@{}}
$(1.290\times10^{-5},\;2.804\times10^{-4})$\\
$(-0.162,\;0.033)$
\end{tabular}
&
$(0.965,\;1.067,\;1.150)$
&
\begin{tabular}{@{}c@{}}
$(2.530,\;1.100)$\\
$(1.200,\;1.500)$
\end{tabular}
\\[3mm]

&
\textbf{Syn}
&
\begin{tabular}{@{}c@{}}
$(35.040,\;495.678,$\\
$1298.011)$
\end{tabular}
&
$(1.786,\;5.859)$
&
\begin{tabular}{@{}c@{}}
$(1.327\times10^{-5},\;2.805\times10^{-4})$\\
$(-0.154,\;0.032)$
\end{tabular}
&
$(0.975,\;1.087,\;1.139)$
&
\\[3mm]

&
\textbf{MM}
&
\begin{tabular}{@{}c@{}}
$(35.002,\;498.712,$\\
$1251.210)$
\end{tabular}
&
$(1.695,\;5.865)$
&
\begin{tabular}{@{}c@{}}
$(1.253\times10^{-5},\;2.802\times10^{-4})$\\
$(-0.160,\;0.032)$
\end{tabular}
&
$(0.966,\;1.067,\;1.151)$
&
\\[3mm]

\midrule

\multirow{3}{*}{\textbf{SKA}}
&
\textbf{Fit}
&
\begin{tabular}{@{}c@{}}
$(38.000,\;600.000,$\\
$1200.000)$
\end{tabular}
&
$(1.764,\;5.194)$
&
\begin{tabular}{@{}c@{}}
$(2.056\times10^{-5},\;-3.825\times10^{-4})$\\
$(-0.246,\;0.028)$
\end{tabular}
&
$(0.947,\;1.062,\;1.147)$
&
\begin{tabular}{@{}c@{}}
$(2.450,\;1.300)$\\
$(1.100,\;1.500)$
\end{tabular}
\\[3mm]

&
\textbf{Syn}
&
\begin{tabular}{@{}c@{}}
$(37.977,\;591.206,$\\
$1259.722)$
\end{tabular}
&
$(1.814,\;5.191)$
&
\begin{tabular}{@{}c@{}}
$(2.096\times10^{-5},\;-3.819\times10^{-4})$\\
$(-0.234,\;0.027)$
\end{tabular}
&
$(0.960,\;1.082,\;1.138)$
&
\\[3mm]

&
\textbf{MM}
&
\begin{tabular}{@{}c@{}}
$(37.988,\;599.222,$\\
$1200.649)$
\end{tabular}
&
$(1.638,\;5.202)$
&
\begin{tabular}{@{}c@{}}
$(1.956\times10^{-5},\;-3.821\times10^{-4})$\\
$(-0.245,\;0.027)$
\end{tabular}
&
$(0.948,\;1.062,\;1.151)$
&
\\[3mm]

\midrule

\multirow{3}{*}{\textbf{SKB}}
&
\textbf{Fit}
&
\begin{tabular}{@{}c@{}}
$(65.000,\;500.000,$\\
$1250.000)$
\end{tabular}
&
$(0.125,\;7.281)$
&
\begin{tabular}{@{}c@{}}
$(1.109\times10^{-5},\;-3.297\times10^{-3})$\\
$(-0.186,\;0.083)$
\end{tabular}
&
$(0.963,\;1.064,\;1.147)$
&
\begin{tabular}{@{}c@{}}
$(2.550,\;1.100)$\\
$(1.300,\;1.450)$
\end{tabular}
\\[3mm]

&
\textbf{Syn}
&
\begin{tabular}{@{}c@{}}
$(65.196,\;495.759,$\\
$1296.061)$
\end{tabular}
&
$(0.138,\;7.279)$
&
\begin{tabular}{@{}c@{}}
$(1.143\times10^{-5},\;-3.286\times10^{-3})$\\
$(-0.178,\;0.080)$
\end{tabular}
&
$(0.973,\;1.084,\;1.136)$
&
\\[3mm]

&
\textbf{MM}
&
\begin{tabular}{@{}c@{}}
$(65.385,\;498.421,$\\
$1252.594)$
\end{tabular}
&
$(0.138,\;7.268)$
&
\begin{tabular}{@{}c@{}}
$(1.132\times10^{-5},\;-3.264\times10^{-3})$\\
$(-0.184,\;0.082)$
\end{tabular}
&
$(0.963,\;1.062,\;1.148)$
&
\\[3mm]

\midrule

\multirow{3}{*}{\textbf{SKI2}}
&
\textbf{Fit}
&
\begin{tabular}{@{}c@{}}
$(40.000,\;500.000,$\\
$1250.000)$
\end{tabular}
&
$(0.267,\;8.418)$
&
\begin{tabular}{@{}c@{}}
$(7.407\times10^{-5},\;-3.861\times10^{-3})$\\
$(-0.131,\;0.028)$
\end{tabular}
&
$(0.860,\;0.961,\;1.047)$
&
\begin{tabular}{@{}c@{}}
$(2.250,\;1.100)$\\
$(1.200,\;1.500)$
\end{tabular}
\\[3mm]

&
\textbf{Syn}
&
\begin{tabular}{@{}c@{}}
$(40.058,\;497.760,$\\
$1287.916)$
\end{tabular}
&
$(0.295,\;8.424)$
&
\begin{tabular}{@{}c@{}}
$(7.572\times10^{-5},\;-3.842\times10^{-3})$\\
$(-0.125,\;0.027)$
\end{tabular}
&
$(0.868,\;0.978,\;1.038)$
&
\\[3mm]

&
\textbf{MM}
&
\begin{tabular}{@{}c@{}}
$(40.412,\;499.798,$\\
$1248.356)$
\end{tabular}
&
$(0.236,\;8.444)$
&
\begin{tabular}{@{}c@{}}
$(7.163\times10^{-5},\;-3.858\times10^{-3})$\\
$(-0.131,\;0.028)$
\end{tabular}
&
$(0.862,\;0.962,\;1.049)$
&
\\[3mm]

\midrule

\multirow{3}{*}{\textbf{SKI3}}
&
\textbf{Fit}
&
\begin{tabular}{@{}c@{}}
$(30.000,\;500.000,$\\
$1200.000)$
\end{tabular}
&
$(3.476,\;2.825)$
&
\begin{tabular}{@{}c@{}}
$(3.058\times10^{-5},\;-1.713\times10^{-3})$\\
$(-0.124,\;0.034)$
\end{tabular}
&
$(0.919,\;1.036,\;1.126)$
&
\begin{tabular}{@{}c@{}}
$(2.400,\;1.050)$\\
$(1.250,\;1.500)$
\end{tabular}
\\[3mm]

&
\textbf{Syn}
&
\begin{tabular}{@{}c@{}}
$(30.005,\;495.069,$\\
$1257.847)$
\end{tabular}
&
$(3.698,\;2.822)$
&
\begin{tabular}{@{}c@{}}
$(3.139\times10^{-5},\;-1.712\times10^{-3})$\\
$(-0.118,\;0.033)$
\end{tabular}
&
$(0.932,\;1.057,\;1.118)$
&
\\[3mm]

&
\textbf{MM}
&
\begin{tabular}{@{}c@{}}
$(29.906,\;498.400,$\\
$1200.705)$
\end{tabular}
&
$(3.056,\;2.821)$
&
\begin{tabular}{@{}c@{}}
$(2.912\times10^{-5},\;-1.719\times10^{-3})$\\
$(-0.123,\;0.034)$
\end{tabular}
&
$(0.919,\;1.036,\;1.131)$
&
\\[3mm]

\midrule

\multirow{3}{*}{\textbf{SKI4}}
&
\textbf{Fit}
&
\begin{tabular}{@{}c@{}}
$(30.000,\;500.000,$\\
$1100.000)$
\end{tabular}
&
$(8.129,\;1.555)$
&
\begin{tabular}{@{}c@{}}
$(5.705\times10^{-6},\;4.135\times10^{-5})$\\
$(0.167,\;-0.286)$
\end{tabular}
&
$(0.866,\;0.994,\;1.085)$
&
\begin{tabular}{@{}c@{}}
$(2.650,\;1.050)$\\
$(1.400,\;1.300)$
\end{tabular}
\\[3mm]

&
\textbf{Syn}
&
\begin{tabular}{@{}c@{}}
$(29.967,\;495.580,$\\
$1164.626)$
\end{tabular}
&
$(8.446,\;1.552)$
&
\begin{tabular}{@{}c@{}}
$(5.932\times10^{-6},\;4.137\times10^{-5})$\\
$(0.159,\;-0.270)$
\end{tabular}
&
$(0.882,\;1.013,\;1.080)$
&
\\[3mm]

&
\textbf{MM}
&
\begin{tabular}{@{}c@{}}
$(29.998,\;496.664,$\\
$1100.567)$
\end{tabular}
&
$(8.319,\;1.557)$
&
\begin{tabular}{@{}c@{}}
$(5.856\times10^{-6},\;4.129\times10^{-5})$\\
$(0.165,\;-0.281)$
\end{tabular}
&
$(0.865,\;0.997,\;1.085)$
&
\\[3mm]

\midrule

\multirow{3}{*}{\textbf{SKI6}}
&
\textbf{Fit}
&
\begin{tabular}{@{}c@{}}
$(35.000,\;500.000,$\\
$1400.000)$
\end{tabular}
&
$(6.586,\;2.047)$
&
\begin{tabular}{@{}c@{}}
$(5.867\times10^{-6},\;-1.184\times10^{-5})$\\
$(0.069,\;-0.356)$
\end{tabular}
&
$(0.962,\;1.047,\;1.116)$
&
\begin{tabular}{@{}c@{}}
$(2.650,\;1.600)$\\
$(1.400,\;1.100)$
\end{tabular}
\\[3mm]

&
\textbf{Syn}
&
\begin{tabular}{@{}c@{}}
$(35.044,\;493.264,$\\
$1390.687)$
\end{tabular}
&
$(6.722,\;2.049)$
&
\begin{tabular}{@{}c@{}}
$(5.980\times10^{-6},\;-1.183\times10^{-5})$\\
$(0.068,\;-0.346)$
\end{tabular}
&
$(0.961,\;1.062,\;1.099)$
&
\\[3mm]

&
\textbf{MM}
&
\begin{tabular}{@{}c@{}}
$(34.984,\;496.636,$\\
$1398.394)$
\end{tabular}
&
$(6.619,\;2.047)$
&
\begin{tabular}{@{}c@{}}
$(5.896\times10^{-6},\;-1.184\times10^{-5})$\\
$(0.068,\;-0.348)$
\end{tabular}
&
$(0.966,\;1.048,\;1.118)$
&
\\[3mm]

\midrule

\multirow{3}{*}{\textbf{SKMP}}
&
\textbf{Fit}
&
\begin{tabular}{@{}c@{}}
$(60.000,\;500.000,$\\
$1400.000)$
\end{tabular}
&
$(1.009,\;5.386)$
&
\begin{tabular}{@{}c@{}}
$(2.756\times10^{-5},\;-4.646\times10^{-4})$\\
$(-0.565,\;0.431)$
\end{tabular}
&
$(0.930,\;1.020,\;1.094)$
&
\begin{tabular}{@{}c@{}}
$(2.400,\;1.550)$\\
$(1.300,\;1.350)$
\end{tabular}
\\[3mm]

&
\textbf{Syn}
&
\begin{tabular}{@{}c@{}}
$(60.089,\;497.184,$\\
$1373.846)$
\end{tabular}
&
$(1.049,\;5.386)$
&
\begin{tabular}{@{}c@{}}
$(2.809\times10^{-5},\;-4.645\times10^{-4})$\\
$(-0.550,\;0.421)$
\end{tabular}
&
$(0.926,\;1.032,\;1.078)$
&
\\[3mm]

&
\textbf{MM}
&
\begin{tabular}{@{}c@{}}
$(60.102,\;502.760,$\\
$1395.618)$
\end{tabular}
&
$(1.036,\;5.387)$
&
\begin{tabular}{@{}c@{}}
$(2.792\times10^{-5},\;-4.643\times10^{-4})$\\
$(-0.566,\;0.432)$
\end{tabular}
&
$(0.929,\;1.022,\;1.092)$
&
\\[3mm]

\midrule

\multirow{3}{*}{\textbf{SKOP}}
&
\textbf{Fit}
&
\begin{tabular}{@{}c@{}}
$(60.000,\;600.000,$\\
$1400.000)$
\end{tabular}
&
$(2.153,\;4.441)$
&
\begin{tabular}{@{}c@{}}
$(2.188\times10^{-5},\;-1.300\times10^{-3})$\\
$(0.057,\;-0.300)$
\end{tabular}
&
$(0.815,\;0.894,\;0.959)$
&
\begin{tabular}{@{}c@{}}
$(2.400,\;1.100)$\\
$(1.400,\;1.100)$
\end{tabular}
\\[3mm]

&
\textbf{Syn}
&
\begin{tabular}{@{}c@{}}
$(60.020,\;595.759,$\\
$1355.185)$
\end{tabular}
&
$(2.210,\;4.440)$
&
\begin{tabular}{@{}c@{}}
$(2.227\times10^{-5},\;-1.299\times10^{-3})$\\
$(0.055,\;-0.288)$
\end{tabular}
&
$(0.809,\;0.902,\;0.946)$
&
\\[3mm]

&
\textbf{MM}
&
\begin{tabular}{@{}c@{}}
$(59.982,\;599.296,$\\
$1394.561)$
\end{tabular}
&
$(2.267,\;4.435)$
&
\begin{tabular}{@{}c@{}}
$(2.264\times10^{-5},\;-1.299\times10^{-3})$\\
$(0.058,\;-0.301)$
\end{tabular}
&
$(0.814,\;0.895,\;0.958)$
&
\\[3mm]

\midrule


\multirow{3}{*}{\textbf{SLY}}
&
\textbf{Fit}
&
\begin{tabular}{@{}c@{}}
$(200.000,\;600.000,$\\
$1200.000)$
\end{tabular}
&
$(14.993,\;1.775)$
&
\begin{tabular}{@{}c@{}}
$(-6.441\times10^{-4},\;1.754\times10^{-4})$\\
$(-0.433,\;0.328)$
\end{tabular}
&
$(0.903,\;1.017,\;1.092)$
&
\begin{tabular}{@{}c@{}}
$(1.900,\;2.200)$\\
$(1.350,\;1.400)$
\end{tabular}
\\[3mm]

&
\textbf{Syn}
&
\begin{tabular}{@{}c@{}}
$(199.701,\;597.603,$\\
$1257.330)$
\end{tabular}
&
$(15.728,\;1.776)$
&
\begin{tabular}{@{}c@{}}
$(-6.396\times10^{-4},\;1.767\times10^{-4})$\\
$(-0.416,\;0.316)$
\end{tabular}
&
$(0.916,\;1.036,\;1.083)$
&
\\[3mm]

&
\textbf{MM}
&
\begin{tabular}{@{}c@{}}
$(200.063,\;601.419,$\\
$1201.885)$
\end{tabular}
&
$(14.249,\;1.774)$
&
\begin{tabular}{@{}c@{}}
$(-6.415\times10^{-4},\;1.726\times10^{-4})$\\
$(-0.431,\;0.326)$
\end{tabular}
&
$(0.903,\;1.018,\;1.093)$
&
\\[3mm]

\midrule

\multirow{3}{*}{\textbf{SLY2}}
&
\textbf{Fit}
&
\begin{tabular}{@{}c@{}}
$(200.000,\;600.000,$\\
$1200.000)$
\end{tabular}
&
$(15.742,\;1.777)$
&
\begin{tabular}{@{}c@{}}
$(-3.299\times10^{-4},\;2.979\times10^{-5})$\\
$(-0.074,\;5.024\times10^{-3})$
\end{tabular}
&
$(0.861,\;0.954,\;1.028)$
&
\begin{tabular}{@{}c@{}}
$(1.800,\;2.400)$\\
$(1.200,\;1.700)$
\end{tabular}
\\[3mm]

&
\textbf{Syn}
&
\begin{tabular}{@{}c@{}}
$(200.124,\;592.221,$\\
$1252.241)$
\end{tabular}
&
$(16.224,\;1.779)$
&
\begin{tabular}{@{}c@{}}
$(-3.283\times10^{-4},\;3.027\times10^{-5})$\\
$(-0.069,\;4.849\times10^{-3})$
\end{tabular}
&
$(0.869,\;0.971,\;1.020)$
&
\\[3mm]

&
\textbf{MM}
&
\begin{tabular}{@{}c@{}}
$(199.725,\;599.562,$\\
$1198.463)$
\end{tabular}
&
$(14.840,\;1.776)$
&
\begin{tabular}{@{}c@{}}
$(-3.303\times10^{-4},\;2.883\times10^{-5})$\\
$(-0.074,\;4.936\times10^{-3})$
\end{tabular}
&
$(0.862,\;0.955,\;1.027)$
&
\\[3mm]

\midrule

\multirow{3}{*}{\textbf{SLY4}}
&
\textbf{Fit}
&
\begin{tabular}{@{}c@{}}
$(200.000,\;600.000,$\\
$1300.000)$
\end{tabular}
&
$(14.993,\;1.775)$
&
\begin{tabular}{@{}c@{}}
$(-6.024\times10^{-4},\;3.986\times10^{-4})$\\
$(-0.302,\;0.173)$
\end{tabular}
&
$(0.919,\;1.029,\;1.101)$
&
\begin{tabular}{@{}c@{}}
$(2.100,\;2.200)$\\
$(1.300,\;1.400)$
\end{tabular}
\\[3mm]

&
\textbf{Syn}
&
\begin{tabular}{@{}c@{}}
$(200.230,\;595.449,$\\
$1323.240)$
\end{tabular}
&
$(15.641,\;1.775)$
&
\begin{tabular}{@{}c@{}}
$(-6.002\times10^{-4},\;3.992\times10^{-4})$\\
$(-0.291,\;0.167)$
\end{tabular}
&
$(0.925,\;1.046,\;1.089)$
&
\\[3mm]

&
\textbf{MM}
&
\begin{tabular}{@{}c@{}}
$(200.244,\;600.138,$\\
$1298.006)$
\end{tabular}
&
$(15.529,\;1.775)$
&
\begin{tabular}{@{}c@{}}
$(-5.986\times10^{-4},\;3.979\times10^{-4})$\\
$(-0.301,\;0.173)$
\end{tabular}
&
$(0.919,\;1.030,\;1.101)$
&
\\[3mm]

\midrule

\multirow{3}{*}{\textbf{WFF1}}
&
\textbf{Fit}
&
\begin{tabular}{@{}c@{}}
$(30.000,\;600.000,$\\
$1000.000)$
\end{tabular}
&
$(3.067,\;4.727)$
&
\begin{tabular}{@{}c@{}}
$(1.574\times10^{-8},\;1.581\times10^{-3})$\\
$(-0.113,\;0.010)$
\end{tabular}
&
$(1.150,\;1.405,\;1.484)$
&
\begin{tabular}{@{}c@{}}
$(3.500,\;1.200)$\\
$(1.300,\;1.700)$
\end{tabular}
\\[3mm]

&
\textbf{Syn}
&
\begin{tabular}{@{}c@{}}
$(29.939,\;605.586,$\\
$998.433)$
\end{tabular}
&
$(3.075,\;4.729)$
&
\begin{tabular}{@{}c@{}}
$(1.613\times10^{-8},\;1.583\times10^{-3})$\\
$(-0.108,\;0.010)$
\end{tabular}
&
$(1.148,\;1.415,\;1.480)$
&
\\[3mm]

&
\textbf{MM}
&
\begin{tabular}{@{}c@{}}
$(29.929,\;589.567,$\\
$1001.618)$
\end{tabular}
&
$(3.106,\;4.716)$
&
\begin{tabular}{@{}c@{}}
$(1.624\times10^{-8},\;1.591\times10^{-3})$\\
$(-0.114,\;0.011)$
\end{tabular}
&
$(1.143,\;1.405,\;1.484)$
&
\\[3mm]

\midrule

\multirow{3}{*}{\textbf{WFF2}}
&
\textbf{Fit}
&
\begin{tabular}{@{}c@{}}
$(50.000,\;500.000,$\\
$950.000)$
\end{tabular}
&
$(6.849,\;2.789)$
&
\begin{tabular}{@{}c@{}}
$(3.827\times10^{-8},\;7.729\times10^{-4})$\\
$(-1.068\times10^{-3},\;6.454\times10^{-4})$
\end{tabular}
&
$(1.205,\;1.298,\;1.378)$
&
\begin{tabular}{@{}c@{}}
$(3.400,\;1.400)$\\
$(2.100,\;2.200)$
\end{tabular}
\\[3mm]

&
\textbf{Syn}
&
\begin{tabular}{@{}c@{}}
$(50.052,\;494.101,$\\
$930.504)$
\end{tabular}
&
$(6.840,\;2.793)$
&
\begin{tabular}{@{}c@{}}
$(3.851\times10^{-8},\;7.736\times10^{-4})$\\
$(-1.035\times10^{-3},\;6.279\times10^{-4})$
\end{tabular}
&
$(1.194,\;1.308,\;1.374)$
&
\\[3mm]

&
\textbf{MM}
&
\begin{tabular}{@{}c@{}}
$(50.060,\;506.093,$\\
$954.470)$
\end{tabular}
&
$(6.927,\;2.785)$
&
\begin{tabular}{@{}c@{}}
$(3.983\times10^{-8},\;7.768\times10^{-4})$\\
$(-1.062\times10^{-3},\;6.453\times10^{-4})$
\end{tabular}
&
$(1.192,\;1.299,\;1.374)$
&
\\[3mm]

\midrule

\multirow{3}{*}{\textbf{WFF3}}
&
\textbf{Fit}
&
\begin{tabular}{@{}c@{}}
$(50.000,\;500.000,$\\
$1100.000)$
\end{tabular}
&
$(6.849,\;2.789)$
&
\begin{tabular}{@{}c@{}}
$(3.025\times10^{-7},\;1.479\times10^{-3})$\\
$(8.158\times10^{-3},\;-0.124)$
\end{tabular}
&
$(0.604,\;0.672,\;0.705)$
&
\begin{tabular}{@{}c@{}}
$(3.050,\;1.200)$\\
$(1.600,\;1.100)$
\end{tabular}
\\[3mm]

&
\textbf{Syn}
&
\begin{tabular}{@{}c@{}}
$(49.986,\;498.646,$\\
$1168.269)$
\end{tabular}
&
$(6.971,\;2.783)$
&
\begin{tabular}{@{}c@{}}
$(3.186\times10^{-7},\;1.483\times10^{-3})$\\
$(7.774\times10^{-3},\;-0.113)$
\end{tabular}
&
$(0.612,\;0.684,\;0.701)$
&
\\[3mm]

&
\textbf{MM}
&
\begin{tabular}{@{}c@{}}
$(49.998,\;496.058,$\\
$1104.062)$
\end{tabular}
&
$(6.843,\;2.788)$
&
\begin{tabular}{@{}c@{}}
$(3.001\times10^{-7},\;1.480\times10^{-3})$\\
$(7.885\times10^{-3},\;-0.119)$
\end{tabular}
&
$(0.598,\;0.672,\;0.704)$
&
\\[3mm]

\midrule

\end{longtable}

\twocolumn

\def\bibfont{\small}
\bibliography{ref.bib}
\end{document}